\documentstyle [10pt,eqsecnum,aps,amsfonts] {revtex}
\input epsf
\newcommand{\beq}{\begin{equation}}
\newcommand{\eeq}{\end{equation}}
\newcommand{\beqs}{\begin{eqnarray}}
\newcommand{\eeqs}{\end{eqnarray}}

\begin{document}
\draft

\baselineskip 5.0mm

\title{Ground State Degeneracy of Potts Antiferromagnets: Homeomorphic 
Classes with Noncompact $W$ Boundaries}

\author{Robert Shrock\thanks{email: shrock@insti.physics.sunysb.edu}
\and Shan-Ho Tsai\thanks{current address: Department of Physics and Astronomy,
University of Georgia, Athens, GA 30602; email: tsai@hal.physast.uga.edu}}

\address{
Institute for Theoretical Physics  \\
State University of New York       \\
Stony Brook, N. Y. 11794-3840}

\maketitle

\vspace{10mm}

\begin{abstract}

We present exact calculations of the zero-temperature partition function 
$Z(G,q,T=0)$ and ground-state degeneracy $W(\{G\},q)$ for the $q$-state 
Potts antiferromagnet on a number of families of graphs $G$ for which 
(generalizing $q$ from ${\mathbb Z}_+$ to ${\mathbb C}$) the boundary 
${\cal B}$ of regions of analyticity of $W$ in the complex $q$ plane is 
noncompact, passing through $z=1/q=0$.   For these types of graphs, since 
the reduced function $W_{red.}=q^{-1}W$ is nonanalytic at $z=0$, there 
is no large--$q$ Taylor series expansion of $W_{red.}$.  The study of these
graphs thus gives insight into the conditions for the validity of the
large--$q$ expansions.  It is shown how such (families of) graphs can be 
generated from known families by homeomorphic expansion.  

\end{abstract}


\vspace{16mm}

\pagestyle{empty}
\newpage

\pagestyle{plain}
\pagenumbering{arabic}
\renewcommand{\thefootnote}{\arabic{footnote}}
\setcounter{footnote}{0}

\section{Introduction}

    Nonzero ground state entropy, $S_0 \ne 0$, is an important subject in 
statistical mechanics.  One physical example is provided by ice, for which 
$S_0 = 0.82 \pm 0.05$ cal/(K-mole), i.e., $S_0/R = 0.41 \pm 0.03$, where
$R=N_{Avog.}k_B$ \cite{ice,lp,liebwu}. A particularly simple model exhibiting
ground state entropy without the complication of frustration is the $q$-state
Potts antiferromagnet (AF) \cite{potts,fk,wurev} on a graph $G$ 
\cite{graphdef} 
(which may or may not be a regular lattice) for $q \ge \chi(G)$, where 
$\chi(\Lambda)$ denotes the chromatic number of $G$, i.e., the minimum 
number of colors necessary to color the vertices of the lattice such that no 
two adjacent vertices have the same color.  As is evident, there is a deep 
connection with graph theory here, since the zero-temperature partition 
function of the above-mentioned $q$-state Potts antiferromagnet on a graph 
$G$ satisfies
\beq
Z(G,q,T=0)_{PAF}=P(G,q)
\label{zp}
\eeq
where $P(G,q)$ is the chromatic
polynomial expressing the number of ways of coloring the vertices of the graph
$G$ with $q$ colors such that no two adjacent vertices (connected by a bond of
the graph) have the same color \cite{birk}--\cite{biggsbook}.
Since the strict mathematical definition of a graph $G$ requires the number of
vertices $n=v(G)$ to be finite, we shall denote
\beq
\lim_{n \to \infty}G = \{G\}
\label{ginfinite}
\eeq
In this limit, the ground state entropy per vertex (site) is given by 
\beq
S_0 = k_B \ln W(\{G\},q)
\label{s0}
\eeq
where $W(\{G\},q)$, the ground state degeneracy per vertex, is
\beq
W(\{G\},q) = \lim_{n \to \infty} P(G,q)^{1/n}
\label{w}
\eeq
As noted, in the limit $n \to \infty$, nonzero ground state entropy, 
$S_0(\{G\},q) > 0$, or equivalently, ground state degeneracy $W(\{G\},q) > 1$ 
generically occurs in the $q$-state Potts antiferromagnet for sufficiently 
large $q$ on a given lattice $G$; equivalently, the number of ways of coloring
the graph subject to the constraint that no two adjacent vertices have the
same color grows exponentially rapidly with the number of vertices on the
graph.  

Given the fact that $P(G,q)$ is a polynomial, there is a natural
generalization, which we assume here, of the variable $q$ from integer to
complex values.  $W(\{G\},q)$ is an analytic function in the $q$ plane except
along a continuous locus of points (and at possible isolated points which will
not be relevant here).
Following the terminology in our earlier papers \cite{p3afhc}--\cite{strip}, we
denote this continuous locus as ${\cal B}$. In the limit as $n \to \infty$, the
locus ${\cal B}$ forms by means of a coalescence of a subset of the zeros of
$P(G,q)$ (called chromatic zeros of $G$) \cite{early}.  Because $P(G,q)$ has
real (indeed, integer) coefficients, ${\cal B}$ has the basic property of
remaining invariant under the replacement $q \to q^*$: 
\beq 
{\cal B}(q) = {\cal B}(q^*)
\label{binvariance}
\eeq
In a 
series of papers we have calculated and analyzed the loci ${\cal B}$ for a 
variety of families of graphs \cite{p3afhc}--\cite{strip} and investigated the
connections of these loci with the behavior of $W(\{G\},q)$ for the physical
values $q \in {\mathbb Z}_+$, which we have also studied by means of rigorous
upper and lower bounds and large--$q$ series \cite{ww,wn} and Monte Carlo 
measurements \cite{p3afhc,w,ww}.  In many cases the loci ${\cal B}$ enclose 
and form the boundaries of two or more regions in the complex $q$ plane. 
Since this is the case for the families of graphs studied here, we shall use
the abbreviated nomenclature ``$W$ boundary'' for ${\cal B}$.  We note that in
other families of graphs, such as the infinitely long, finite-width 
homogeneous strip graphs of various lattices (with open boundary conditions) 
studied in Ref. \cite{strip}, the loci ${\cal B}$ form arcs that do 
not enclose any region.  The study of these loci for various families of 
graphs also makes a very interesting connection with algebraic geometry. 

With regard to the properties of the ground state entropy of the Potts
antiferromagnet on a given ($n \to \infty$ limit of a graph) $\{G\}$, two of 
the most important properties of 
$W(\{G\},q)$ and the associated nonanalytic locus ${\cal B}$ include (i) 
the maximal finite value of $q$ at which ${\cal B}$ crosses the real axis in
the $q$ plane, which we denote $q_c$, as in our previous work; and (ii) the
question of whether ${\cal B}(q)$ is bounded or unbounded (noncompact). 
Property (i) is important because in cases where ${\cal B}$ encloses regions,
the behavior of the ground state degeneracy $W(\{G\},q)$ changes qualitatively
(nonanalytically) as $q$ (taken to be real) decreases through
$q_c$. \cite{exception}. 
Concerning the importance of property (ii), we first recall that 
since an obvious upper bound on $P(G,q)$
describing the coloring of an $n$-vertex graph with $q$ colors is 
$P(G,q) \le q^n$, and hence $W(\{G\},q) \le q$, it is natural to define a 
reduced function that is bounded as $q \to \infty$: 
\beq
W_{red.}(\{G\},q) = q^{-1}W(\{G\},q)
\label{wr}
\eeq
(This function was denoted $W_r(\{G\},q)$ in our Refs. \cite{w,wa}.) 
If and only if $W_{red.}(\{G\},q)$ is analytic at $z=1/q=0$, there exists a 
Taylor
series expansion of this function around this point.  Such large--$q$ series
expansions are a useful means for obtaining approximate values of the 
ground state degeneracy of the $q$-state Potts antiferromagnet for finite $q$
on regular lattices \cite{nagle,baker,kewser,ww,wn}.  In turn, 
$W_{red.}(\{G\},q)$ 
is nonanalytic at $1/q=0$ if and only if the locus ${\cal B}$
is unbounded in the $q$ plane, i.e., passes through the origin of the $1/q$
plane.  Thus, families of graphs with unbounded ${\cal B}(q)$ do not have 
large--$q$ Taylor series expansions for $W_{red.}(\{G\},q)$.  
In Ref. \cite{w} we drew attention to this for bipyramid graphs 
\cite{read91} and in Ref. \cite{wa} 
we presented a general method for constructing families of graphs with
unbounded, noncompact loci ${\cal B}$.  We also presented a general (necessary
and sufficient) algebraic condition such that ${\cal B}$ is unbounded. 
Clearly it is important to understand better the differences between 
the families of graphs that yield $W_{red.}(\{G\},q)$ functions analytic at
$1/q=0$ and those that produce $W_{red.}(\{G\},q)$ functions that are
nonanalytic at $1/q=0$.  Since ${\cal B}$ forms by merging of 
chromatic zeros of $G$ as $v(G) \to \infty$, a necessary condition for 
${\cal B}$ to be noncompact in the $q$ plane, extending infinitely far from the
origin, is that as $v(G) \to \infty$, the magnitudes $|q|$ of some chromatic 
zeros must increase without bound.  That this is not a sufficient condition is
illustrated by the chromatic zeros of the complete graph $K_p$
\cite{complete}: these occur at $q=0,1,..,p-1$ so that, as $p \to \infty$, the
magnitude of the largest chromatic zero grows without bound, but the locus
${\cal B}$ for this family is not only not noncompact, it is trivial,
consisting of the empty set.  

In the present paper, continuing our earlier studies
\cite{w,wa}, we shall address this problem.   We shall present several 
additional methods for constructing families of graphs with unbounded 
${\cal B}$. We shall use these to obtain exact
calculations of the corresponding chromatic polynomials, $W$ functions, and 
boundaries ${\cal B}$.  In particular, we shall generate an infinite number of
such families via the method of homeomorphic expansion.  Homeomorphic classes
of graphs have been of continuing interest in graph theory \cite{biggsbook}; 
for some recent theorems in a somewhat different direction from 
the present work, see Ref. \cite{rw}.  
The study of these families is valuable because it 
enables one gain further insight into the origin of the nonanalyticity of
$W_r(\{G\},q)$ at $1/q=0$ and the consequent nonexistence of large--$q$ series
expansions for the ground state degeneracy of the Potts antiferromagnet on such
graphs.  Here we shall give detailed analyses of families with loci ${\cal B}$ 
that have a simple structure in the vicinity of $1/q=0$.  In a companion 
paper, we shall present results for families of
graphs with loci ${\cal B}$ that are more complicated in the vicinity of
$1/q=0$.

   Before proceeding, we recall two subtleties in the definition of 
$W(\{G\},q)$ when $q$ is not a positive integer \cite{w}.  First, 
for certain ranges of real $q$, $P(G,q)$ can be negative, and, of course,
when $q$ is complex, so is $P(G,q)$ in general; in these cases it is not 
obvious, {\it a priori}, which of the $n$ roots
\beq
P(G,q)^{1/n} = \{ |P(G,q)|^{1/n}e^{i \arg(P(G,q))/n}e^{2\pi i r/n} \} \ , 
\quad r=0,1,...,n-1
\label{pphase}
\eeq
to choose in eq. (\ref{w}).
Consider the function $W(\{G\},q)$ defined via eq. (\ref{w})
starting with $q$ on the positive real axis where $P(G,q) > 0$, and consider
the maximal region in the complex $q$ plane that can be reached by analytic
continuation of this function.  We denote this region as $R_1$.  Clearly, the
phase choice in (\ref{pphase}) for $q \in R_1$ is that given by $r=0$. 
For families of graphs $\{G\}$ where there are regions $R_i$ of analyticity of 
$W(\{G\},q)$ that are not analytically connected with $R_1$, there is no
canonical choice of phase in eq. (\ref{pphase}) and hence it is only 
possible to determine the magnitude $|W(\{G\},q)|$ unambiguously.

   A second subtlety in the definition of $W(\{G\},q)$ concerns the fact that
at certain special points $q_s$, the following limits do not commute \cite{w}
(for any choice of $r$ in eq. (\ref{pphase})):
\beq
\lim_{n \to \infty} \lim_{q \to q_s} P(G,q)^{1/n} \ne
\lim_{q \to q_s} \lim_{n \to \infty} P(G,q)^{1/n}
\label{wnoncomm}
\eeq
The set $q_s$ includes points where $P(G,q)$ has zeros whose multiplicity 
does not scale like the number of vertices \cite{w} (typically these are simple
zeros).  Following our earlier work \cite{w}, we define 
\beq
W(\{G\},q_s) \equiv W(\{G\},q_s)_{D_{qn}} \equiv
\lim_{q \to q_s} \lim_{n \to \infty} P(G,q)^{1/n}
\label{wdefqn}
\eeq
This definition has the advantage of maintaining the analyticity of 
$W(\{G\},q)$ at the special points $q_s$.

   A general form for the chromatic polynomial of an $n$-vertex graph $G$ is
\beq
P(G_n,q) =  c_0(q) + \sum_{j=1}^{N_a} c_j(q)a_j(q)^{t(n)} 
\label{pgsum}
\eeq
where
\beq
t(n) = t_1 n + t_0
\label{tn}
\eeq
and $c_j(q)$ and $a_j(q)$ are certain functions of $q$. Here the $a_j(q)$ and
$c_{j \ne 0}(q)$ are independent of $n$, while $c_0(q)$ may contain
$n$-dependent terms, such as $(-1)^n$, but does not grow with $n$ like 
$(const.)^n$.
Obviously, the reality of $P(G,q)$ for real $q$ implies that $c_j(q)$ and
$a_j(q)$ are real for real $q$.  The condition that ${\cal B}$ does not 
extend to infinite distance from the origin in the $q$ plane is
equivalent to the condition that for sufficiently large $|q|$, there is one
leading term $a_j(q)$ in eq. (\ref{pgsum}).  Here we recall that ``leading
term $a_\ell(q)$ at a point $q$'' was defined in Ref. \cite{w} as a term
satisfying $|a_\ell(q)| \ge 1$ and $|a_\ell(q)| > |a_j(q)|$ for $j \ne \ell$.
(If the $c_0$ term is absent and $N_a=1$, then the sole $a_1(q)$ may be 
considered to be leading even if $|a_j(q)| < 1$.)  In the limit as $n \to
\infty$ (i.e., for fixed $p$, $r \to \infty$), the leading term $a_\ell$ in a 
given region determines the limiting function $W$, with $|W|=|a_\ell|^{t_1}$ 
and the boundary ${\cal B}$ occurs where there is a nonanalytic change in $W$ 
as it switches between being determined by different leading terms $a_\ell$ 
in eq. (\ref{pgsum}).

   Since for the families of graphs studied here, the boundary ${\cal B}$ 
is noncompact in the $q$ plane, it is often more convenient to describe the 
boundary in the complex $z$ or $y$ planes, where
\beq
z \equiv \frac{1}{q}
\label{z}
\eeq
and
\beq
y \equiv \frac{1}{q-1}
\label{y}
\eeq 
The variable $y$ is commonly used in large--$q$ series expansions.  Some useful
relations between these variables are 
\beq
q = 1 + \frac{1}{y} \ , \qquad z = \frac{y}{1+y} \ , \qquad y = \frac{z}{1-z}
\label{zyrel}
\eeq
We define polar coordinates as 
\beq
z = \zeta e^{i\theta}
\label{zpolar}
\eeq
and
\beq
y = \rho e^{i\beta}
\label{ypolar}
\eeq

We shall 
use some standard notation from graph theory and combinatorics: 
$v(G)$ and $e(G)$ denote the numbers of vertices and edges in the graph $G$,
and the symbol $q^{(n)}$ is given by 
\beq
q^{(p)} = p! {q \choose q-p} = \prod_{j=0}^{p-1}(q-s)
\label{ff}
\eeq
The terms ``edge'' and ``bond'' will be used synonymously. 
For any partition of $n$, $n=p+u$, 
\beq
q^{(p)}(q-p)^{(u)} = q^{(u)}(q-u)^{(p)} = q^{(p+u)}
\label{ffidentity}
\eeq
For our results below, it will also be convenient to define the polynomial 
\beq
D_k(q) = \frac{P(C_k,q)}{q(q-1)} = a^{k-2}\sum_{j=0}^{k-2}(-a)^{-j} =
\sum_{s=0}^{k-2}(-1)^s {{k-1}\choose {s}} q^{k-2-s}
\label{dk}
\eeq
where
\beq
a=q-1
\label{a}
\eeq
and $P(C_k,q)$ is the chromatic polynomial for the circuit graph,
\beq
P(C_k,q) = a^k + (-1)^ka 
\label{pck}
\eeq
Some useful properties of $D_k$ are listed in an appendix.

The organization of the paper is as follows.  In Section II we review some
methods and results from our previous work on families of the form 
$(K_p)_b + G_r$ that will be necessary for our current discussion (where $K_p$
is the complete graph on $p$ vertices \cite{complete} and, 
following standard notation in the mathematical literature on graph theory, we
use the symbol $G+H$ for the ``join'' of these graphs; see below for
definitions).  Section III contains another method of constructing 
families of graphs with noncompact $W$
boundaries ${\cal B}$ involving the removal of bonds not just from one vertex
of the $K_p$ subgraph in $K_p + G_r$.  In Section IV we present and discuss a
number of ways to construct families of graphs with noncompact $W$ boundaries 
based on homeomorphic expansion of starting sets of graphs.  In Section V we
analyze two such methods involving homeomorphic expansion of the $K_p$ subgraph
in a larger graph.  Sections VI-VIII are detailed analyses of the respective
chromatic polynomials, $W$ functions, and their boundaries of regions of 
analyticity ${\cal B}$ for families of graphs obtained by specific 
homeomorphic expansions.  In Section IX we include some remarks on general
geometrical features of the families of graphs with noncompact $W$ boundaries
${\cal B}$.  Finally, Section X contains our conclusions.

\section{Families of the Form 
$(K_{\lowercase{p}})_{\lowercase{b}}+G_{\lowercase{r}}$ }

Since we will construct and analyze a number of new families of graphs with 
noncompact boundaries ${\cal B}$ as homeomorphic expansions of families 
studied in Ref. \cite{wa}
it is first necessary to review briefly the method that we formulated and
used in Ref. \cite{wa} to generate such families.  Consider a family of 
graphs $G$ and its $v(G) \to \infty$ limit, $\{G\}$ .  
If this family already has the
property that the limiting function $W(\{G\},q)$ has a region boundary
${\cal B}$ that extends to complex infinity in the $q$ plane, then we have no
work to do to get such a boundary. So assume that this family is such that
$W(\{G\},q)$ has a region boundary ${\cal B}$ that does not extend to complex
infinity in the $q$ plane.   Specifically, we start with an
$r$-vertex graph $G_r$ and then form the graph
$K_p + G_r$, where, as above, $K_p$ denotes the complete graph on $p$
vertices \cite{complete}, and we adopt the notation $G + H$ that is commonly
used in graph theory to indicate that every vertex of $G$ is connected by a 
bond to every vertex of $H$; this is called the ``join'' of $G$ and $H$
in the mathematical literature.  We then remove $b$ bonds 
from one vertex in the $K_p$ subgraph; the resultant family is denoted 
\beq
(K_p)_b + G_r
\label{kpbgr}
\eeq
(In Ref. \cite{wa}
 we used the notation $G \times H$ for what is called $G+H$ here,
and the notation $(K_p \times G_n)_{rb}$ for the family denoted 
$(K_p)_b + G_r$ above, where the subscript $rb$ signified ``removed bonds''.)
Since each such vertex has 
degree $\Delta = p-1$ (i.e., has $p-1$ bonds connecting to it), it follows that
\beq
1 \le b \le p-1
\label{bcondition}
\eeq
It was proved in Ref. \cite{wa} that in the
limit $r \to \infty$, the family $(K_p)_b + G_r$ has a noncompact $W$ region 
boundary ${\cal B}(q)$, extending infinitely 
far from the origin of the $q$ plane.  A number of families of this type were
constructed and the chromatic polynomials, resultant $W$ functions, and
boundaries ${\cal B}$ calculated \cite{wa}. Specifically, by means of the 
relation 
\beq
P(K_p + G_r,q) = q^{(p)}P(G_r,q-p) \ , 
\label{pkpg}
\eeq
the chromatic polynomial of $(K_p)_b + G_r$ was calculated to be \cite{wa}
\beqs
P((K_p)_b + G_r,q) =  
P(K_p + G_r,q) + b P(K_{p-1} + G_r,q) \qquad \qquad \qquad \qquad 
\qquad \qquad \cr\cr
= q^{(p-1)}\Bigl [ (q-(p-1))P(G_r,q-p) +b P(G_r,q-(p-1)) \Bigr ]
\label{pkpcutb}
\eeqs
Substituting the expression (\ref{pgsum}) yields 
\beqs
P((K_p)_b + G_r,q) = q^{(p-1)}\biggl [(q-(p-1)) \Big \{ c_0(q-p) +
\sum_{j=1}^{N_a} c_j(q-p)a_j(q-p)^{t(r)} \Big \} \cr
+ b \Big \{ c_0(q-(p-1)) +
\sum_{j=1}^{N_a} c_j(q-(p-1))a_j(q-(p-1))^{t(r)} \Big \} \biggr ]
\label{pkpcutbfull}
\eeqs

For a given $q$, the boundary ${\cal B}$ as $r \to \infty$ (with $p$ fixed) 
is determined by the degeneracy in 
magnitude of the leading terms in eq. (\ref{pkpcutbfull}) at this value of $q$,
viz., 
\beq
|a_\ell(q-p)| = |a_\ell(q-(p-1))|
\label{mageq}
\eeq
This equation, and hence also ${\cal B}$ are independent of $b$ for the
interval (\ref{bcondition}), $1 \le b \le p-1$. In passing, we note that since
the total number of vertices of $(K_p)_b + G_r$ is $v((K_p)_b + G_r)=p+r$,
another way to take $n \to \infty$ is to let $p \to \infty$ with $r$ fixed,
rather than letting $r \to \infty$ with $p$ fixed.  However, from the viewpoint
of either statistical mechanics or graph theory, this is not as interesting a 
limit, since for any given graph $G_r$ and for any given (finite) value of 
$q \in {\mathbb Z}_+$, as $p$ becomes sufficiently large, one will not be able
to color the graph $(K_p)_b + G_r$ and the chromatic polynomial will vanish.
We shall therefore restrict ourselves to the other way of letting the number of
vertices go to infinity, viz., $r \to \infty$ with $p$ fixed. 

Writing 
\beq
a_\ell = \sum_{s=0}^{s_{max}}\alpha_{\ell,s}q^s \ , 
\label{aellsum}
\eeq
recalling that the basic theorem that the coefficient of the
highest-order term, $q^n$, in the chromatic polynomial $P(G_n,q)$ of any
$n$-vertex graph $G_n$ is unity implies that 
\beq
\alpha_{s_{max}}=1 \ , 
\label{alphasmax}
\eeq
next dividing eq. (\ref{mageq}) by $|q^{s_{max}}|$, and finally 
reexpressing the degeneracy equation in the more convenient variable $z$ 
given in eq. (\ref{z}), we obtain
\beqs
& & |(1-pz)^{s_{max}} + 
\sum_{s=0}^{s_{max}-1}\alpha_{\ell,s}(1-pz)^sz^{s_{max}-s}| \cr
& = & 
|(1-(p-1)z)^{s_{max}} + 
\sum_{s=0}^{s_{max}-1}\alpha_{\ell,s}(1-(p-1)z)^sz^{s_{max}-s}|
\label{mageqz}
\eeqs
This equation is clearly satisfied for $z=0$, which shows that the boundary
${\cal B}$ for the $r \to \infty$ limit of the family of graphs 
$(K_p)_b + G_r$ is noncompact in the $q$ plane, passing through $z=1/q=0$.  
As noted above, it is thus more convenient to analyze ${\cal B}$ in the $z$
plane. We next show that ${\cal B}$ passes vertically
through the origin of the $z$ plane for the $r \to \infty$ limit of all of the
families of the type $(K_p)_b + G_r$.  Writing eq. (\ref{mageqz}) in polar
coordinates, using (\ref{zpolar}), we have, for $|z|=\zeta \to 0$, 
\beq
\zeta s_{max} \cos \theta + O(\zeta^2) = 0
\label{rsmax}
\eeq
Since $s_{max} \ne 0$, it follows that
\beq
\cos \theta \to \pm \frac{\pi}{2} \quad {\rm as} \quad \zeta \to 0
\label{vertical}
\eeq
which proves the above assertion, that ${\cal B}$ passes vertically through the
point $z=0$. 

The simplest families of graphs of the form $(K_p)_b + G_r$ are those for which
$P(G_r,q)$ has, in the notation of eq. (\ref{pgsum}), $c_0=0$, $N_a=1$, and 
$a_1(q)$ a linear function of $q$, 
\beq
a_1(q) = q + \alpha_0
\label{a1}
\eeq
For such families, in the $q$ plane, ${\cal B}$ is a vertical line with 
$q_{_I}$ arbitrary and 
\beq
q_{_R} = q_c = p - (\frac{1}{2} + \alpha_0) \quad {\rm for} \quad s_{max}=1
\label{qrsmax1}
\eeq
where $Re(q)=q_{_R}$ and $Im(q)=q_{_I}$. 
In the $z$ plane, ${\cal B}$ is the circle defined by 
\beq
|z-\frac{z_c}{2}| = \frac{z_c}{2}
\label{circlemag}
\eeq
with
\beq
z_c = \frac{1}{q_c} = \frac{1}{p -  (\frac{1}{2} + \alpha_0)}
\label{zcsmax1}
\eeq
Equivalently, ${\cal B}$ is the circle $|y-\frac{y_c}{2}| = \frac{y_c}{2}$ in
the $y$ plane, where $y_c=1/(p-(\frac{3}{2}+\alpha_0))$. 
Thus, ${\cal B}$ divides the $z$ (equivalently $q$ or $y$) plane into two 
regions: (i) $R_1$, as defined above to include the region on the positive 
$z$ axis contiguous with the origin, and (ii) its complement, denoted $R_2$.  
We find that 
\beq
W(\lim_{r \to \infty}[(K_p)_b + G_r],q) = 
q-p+1+\alpha_0 \quad {\rm for} \quad q \in R_1 
\label{wkpgrsmax1region1}
\eeq
\beq
|W(\lim_{r \to \infty}[(K_p)_b + G_r],q)| = 
|q-p+\alpha_0| \quad {\rm for} \quad q \in R_2
\label{wkpgrsmaxregion2}
\eeq
These families include the cases $G_r=$ (i) edgeless graphs
$\overline K_r$; (ii) tree graphs $T_r$; and (iii) chains of triangles with 
each pair of adjacent triangles intersecting on a mutual edge, i.e. 
$(Ch)_{3,r}$ in the notation of Ref. \cite{wa}  (The general chain graph of
$n_p$ $k$-sided polygons with adjacent $k$-gons intersecting on a mutual 
edge is denoted $(Ch)_{k,r}$; the number of vertices, $r$, is given by 
$r=(k-2)n_p+2$.)  The chromatic polynomials for these families are
\beq
P((K_p)_b + \overline K_r ,q) = q^{(p)}\Bigl [ (q-p)^r + b(q-(p-1))^{r-1} 
\Bigr ]
\label{pkper}
\eeq
\beq
P((K_p)_b + T_r,q) = q^{(p+1)}\Bigl [ (q-(p+1))^{r-1} + b(q-p)^{r-2}
\Bigr ]
\label{pkptr}
\eeq
\beq
P((K_p)_b + (Ch)_{k,r},q) = 
q^{(p+1)}\Bigl [ (q-(p+1))D_k(q-p)^{n_p} + bD_k(q-(p-1))^{n_p} \Bigr ]
\label{pkpchainkr}
\eeq
Thus 
\beq
\alpha_0 = 0, \ -1, \ -2 \quad {\rm for} \quad G_r = \overline K_r, \ T_r, \ 
(Ch)_{3,r}
\label{alpha0values}
\eeq
so that 
\beq
q_c = p-\frac{1}{2} \ , \ p+\frac{1}{2} \ , \ p+\frac{3}{2} \ \quad {\rm for}
\quad G_r = \overline K_r \ , \ T_r \ , (Ch)_{3,r}
\label{qcthreecases}
\eeq

\begin{figure}
\centering
\leavevmode
\epsfxsize=3.0in
\begin{center}
\leavevmode
\epsffile{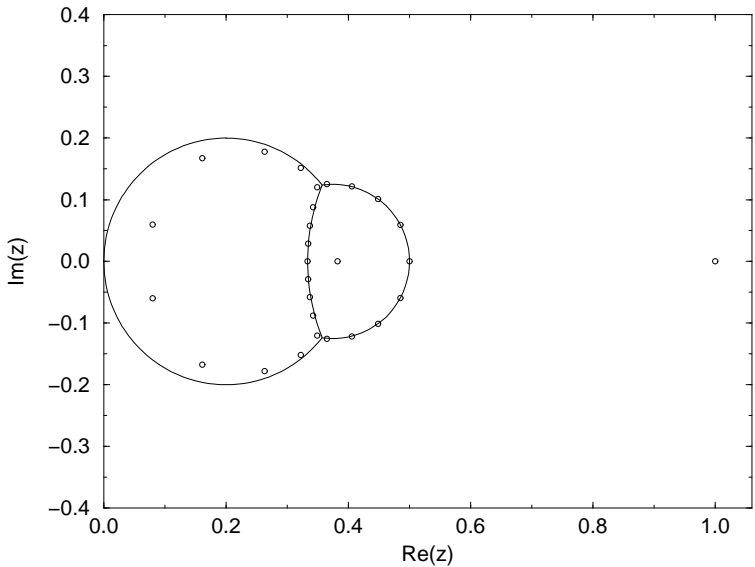}
\end{center}
\vspace{-1cm}
\caption{\footnotesize{Boundary ${\cal B}$ in the $z=1/q$ plane for the $r \to
\infty$ limit of the family of graphs $(K_2)_{b=1} + C_r = \overline K_2 +
C_r$.  Chromatic zeros for $r=27$ are shown for comparison.}}
\label{bipzplane}
\end{figure}

The boundaries ${\cal B}$ for other families have the same behavior near the
origin, $z=0$, as shown by eq. (\ref{rsmax}), but have different features where
they cross the positive real $z$ axis.  For example, consider $G_r=C_r$, 
which has, in the notation of eq. (\ref{pgsum}), $N_a=1$ and a linear 
$a_1(q)=q-1$ but also a nonzero $c_0$ term, viz., $c_0=(-1)^r(q-1)$. 
For $\lim_{r \to \infty}[(K_p)_b + C_r]$, the portion
of ${\cal B}$ near to $z=0$ forms part of the circle (\ref{circlemag}), but to
the right of the points $z_{int.}=1/q_{int.}$ and $z_{int.}^*$, where 
\beq
q_{int.} = p + \frac{1}{2} + \frac{i\sqrt{3}}{2}
\label{qint}
\eeq
the boundary bifurcates into two arcs, which cross the positive $z$ axis at 
\beq
z_c = \frac{1}{q_c} = \frac{1}{p+1} \quad {\rm for} \quad G_r = C_r 
\label{qckpcr}
\eeq
and at $z=1/p$.  Since we shall compare this in detail with homeomorphic
expansions of this family, we show this boundary in Fig. \ref{bipzplane}.  For
this and similar plots given later we also show chromatic zeros calculated for
a reasonably large finite value of $r$ (in this case, $r=27$.  The comparison
of these chromatic zeros with the $r \to \infty$ locus ${\cal B}$ shows
quantitatively how ${\cal B}$ forms via the merging of the zeros.  Of course,
since for finite $r$, the chromatic polynomial is of finite degree (equal to
the number of vertices on the graph), its zeros (i.e., the chromatic zeros of
the graph) are of bounded magnitude in $q$ and hence do not track the
asymptotic curve ${\cal B}$ all the way in to the origin of the $z=1/q$ plane.
The same comment applies to the chromatic zeros that we will show for other
families below. 

Similar comments apply for other families such as 
$(K_p)_b + L_r$ and for the $k \ge 4$ members of the family 
$(K_p)_b + (Ch)_{k,r}$ \cite{wa}, where $L_r$ denotes
a cyclic square strip of width one square, i.e. cyclic ladder graph, and 
$(Ch)_{k,r}$ was defined above in our discussion of the special case $k=3$. 

Of the families of graphs that we have discussed, 
with noncompact boundaries ${\cal B}$ of regions
of analyticity in $W$, most have chromatic polynomials that depend on three 
parameters, $p$, $r$, and $b$, and the resultant ${\cal B}$ for the limit 
$r \to \infty$ depends on one 
parameter, $p$ (and is independent of $b$).  The family 
$(K_p)_b + (Ch)_{k,r}$ illustrates a case where the chromatic
polynomial depends on four parameters -- \ $p$, $k$, $r$, and $b$ -- and 
${\cal B}$ thus depends on two parameters: $p$ and $k$.

\begin{figure}
        \centering
        \leavevmode
\epsfxsize=3.5in
\epsffile{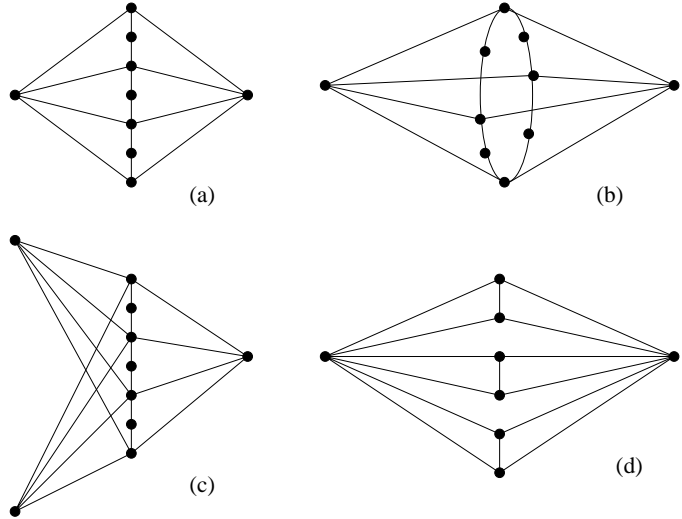}
\caption{\footnotesize{Illustrations of families of graphs considered in the 
text: (a) $T_{k,r} = HEG_{k-2}(\overline K_2 + T_r)$; 
(b) $C_{k,r} = HEG_{k-2} (\overline K_2 + C_r)$; 
(c) $HEG_{k-2}(\overline K_3 + T_r)$; all for $k=3$ and $r=4$;
(d) $\overline K_2 + sK_u$ for $s=3$ and $u=2$. }}
\label{vhegraph1}
\end{figure}

Here we present results on another such family, 
\beq
G_{p,b,s,u} = (K_p)_b + sK_u
\label{gpbsu}
\eeq
where $sK_u$ denotes the disjoint union of 
$s$ copies of the $K_u$ graph (where the vertices of each of the $s$ $K_u$ 
subgraphs are not connected to each other by bonds).  
That is, one adjoins $K_p$ to $G_r = sK_u$ and then removes $b$ bonds from 
one vertex of the $K_p$ subgraph.  An illustration of the special case 
$p=2$, $b=1$, $s=3$, and $u=2$ is shown in Fig. \ref{vhegraph1}(d). 

\vspace{4mm}

Clearly, 
the number of vertices in the disjoint union $sK_u$ is $r=v(sK_u)=su$, so that 
\beq
n=v((K_p)_b + sK_u)=p+su
\label{vkpsku}
\eeq
For $s=1$ and $b=0$, one has the simplification $K_p + K_u = K_{p+u}$.  
Henceforth, we shall assume that $b$ satisfies the restriction
(\ref{bcondition}).  For general $s$ and $u=1$, 
\beq
(K_p)_b + sK_1 = (K_p)_b + \overline K_s
\label{rk1er}
\eeq
so that our analysis for the latter family applies. For general $u$ and $s$, 
from eq. (\ref{pkpcutb}) we have
\beq
P((K_p)_b + sK_u,q) = q^{(p-1)}\Bigl \{ (q-(p-1))[(q-p)^{(u)}]^s 
+ b [(q-(p-1))^{(u)}]^s \Bigr \}
\label{pkprksff}
\eeq
This chromatic polynomial has the form 
\beq
P((K_p)_b + sK_u,q) = q^{(p)}\Bigl [ (q-p)^{(u-1)} \Bigr ]^s P_{b,s}
\label{pkprkub}
\eeq
where $P_{b,s}$ is a polynomial in $q$ of degree $s$.

Now consider the limit $s \to \infty$ with $p$ and $u$ fixed.  Equation 
(\ref{pkprksff}) has the form of eq. (\ref{pgsum}), so that 
the degeneracy equation is 
\beq
|(q-p)^{(u)}| = |(q-(p-1))^{(u)}|
\label{degeqrks}
\eeq
Taking into account that $q=0$ is never a solution of eq. (\ref{degeqrks}) and
dividing by $|q|^u$, we obtain the corresponding degeneracy equation of leading
terms in the $z$ plane:
\beq
|\prod_{j=0}^{u-1}(1-(p+j)z)| = |\prod_{j'=0}^{u-1}(1-(p-1+j')z)|
\label{degeqrksz}
\eeq
By the same reasoning as before, it follows that the resultant locus 
${\cal B}$ which is the solution to the degeneracy equation is noncompact 
in the $q$ plane, passing through $z=1/q=0$.  
The general solution to the degeneracy equation (\ref{degeqrks}) in the $q$
plane is the vertical line given by 
\beq
Re(q) = q_c = p -1 + \frac{u}{2}
\label{bkpku}
\eeq
or equivalently, in the $z$ plane, the circle
\beq
|z - \frac{z_c}{2}| = \frac{z_c}{2} \ , \quad z_c = 
\frac{1}{p-1+\frac{u}{2}}
\label{bkpkuz}
\eeq
and similarly, in the $y$ plane, the circle 
$|y-\frac{y_c}{2}|=\frac{y_c}{2}$ where $y_c = 1/(p-2+\frac{u}{2})$.  
The $q$ (or $z$ or $y$) plane is divided into two regions by the boundary 
${\cal B}$, namely, $R_1$ with $Re(q) > p-1+\frac{u}{2}$, and $R_2$, with 
$Re(q) < p-1+\frac{u}{2}$. 
The dependence of $q_c$ on $u$ is easy to understand: clearly, just as 
increasing $p$ reduces the number of colors available to color the rest of the
graph and hence increases the value of $q_c$ (serving as the demarcation on the
real axis between the region of $q$ values with $W > 1$ and nonzero ground
state entropy $S_0 = k_B \ln W$, from the region where the behavior of $W$ is
different), so also increasing $u$ for fixed $p$ has the same effect.  
We find that 
\beq
W(\lim_{s \to \infty}[(K_p)_b + sK_u],q) = [(q-(p-1))^{(u)}]^{1/u} \quad 
{\rm for} \quad q \in R_1 
\label{wkpkubregion1}
\eeq
where we recall the definition (\ref{ff}).  Further, 
\beq
|W(\lim_{s \to \infty}[(K_p)_b + sK_u],q)| = |(q-p)^{(u)}|^{1/u} 
\quad {\rm for} \quad q \in R_2
\label{wkpkubregion2}
\eeq
It follows that
\beq
W(\lim_{s \to \infty}[(K_p)_b + sK_u],q=q_c) = 0 \quad {\rm if} 
\quad q_c\in {\Bbb Z}_+ \quad {\rm i.e. \ \ if} \quad u \quad {\rm is \ even} 
\label{wboundaryqcz}
\eeq
Note that the zeros in $W$ extend into region $R_1$, to the right of $q_c$, if 
$p+u-2 > q_c$, i.e, if $u \ge 3$.  For example, for the case $u=3$, for which 
$q_c = p + \frac{1}{2}$, the expression (\ref{wkpkubregion1}) has formal 
zeros at $q=p-1$, $q=p$, and $q=p+1$.  The first two of these are not 
relevant since they lie to the left of $q_c$ in the region $R_2$ 
where the expression (\ref{wkpkubregion1}) does not apply; however, the third
zero, at $q=p+1$, is in region $R_1$ and is a true zero of $W$.  For this
illustrative value $u=3$, the expression (\ref{wkpkubregion2}) has 
formal zeros at $q=p$, $q=p+1$, and $q=p+2$.  The last two of these are not
relevant since they lie to the right of $q_c$ in region $R_1$ where the 
expression (\ref{wkpkubregion2}) 
does not apply, while the first one, at $q=p$, lies in region $R_2$
where this expression does apply and hence is a true
zero of $W$ in this region.

It is interesting to comment on the comparative values of $q_c$ defined in the
limit as $r \to \infty$ and the chromatic number $\chi$ for specific members of
a family of graphs parametrized by $r$.  Before doing this, one should note
that there are cases where ${\cal B}$ is trivial or does not cross the real
axis, so that no $q_c$ is defined; an example is given by tree graphs $T_r$,
for which ${\cal B}=\emptyset$ and $\chi=2$.  For circuit graphs $C_r$, 
$\chi=2$ if $r$ is even and $\chi=3$ if $r$ is odd.  In the limit $r \to
\infty$, ${\cal B}$ is the circle $|q-1|=1$, so that $q_c=2$; hence this family
illustrates the possibilities $q_c=\chi$ and $q_c < \chi$.  For the square
lattice with free boundary conditions or (periodic or toroidal) boundary
conditions that preserve the bipartite nature of the lattice, $\chi=2$ while in
all cases $q_c=3$ \cite{w}, so that this illustrates the possibility $q_c >
\chi$.  For the infinitely long strip of the square lattice with width 
$L_y=3$ vertices ($w=L_y-1$ squares), $q_c=2$, equal to $\chi$ for this 
family; for the
infinitely long strip of the triangular lattice with $L_y=3$, $q_c \simeq
2.57$, less than the chromatic number $\chi=3$ for this family, while for the
infinitely long triangular strip with width $L_y=4$, ${\cal B}$ does not cross
the real $q$ axis, so no $q_c$ is defined \cite{w,strip}.  Clearly,
there is a large variety of behavior as regards the question of whether $q_c$
for the infinite-vertex limit of a family of graphs is smaller than, equal to, 
or larger than the chromatic number $\chi$ for specific members of the family.
As an illustration of the families discussed here, 
$\overline K_2 + \overline K_r$ has $\chi=2$ and $q_c=3/2 < \chi$; 
$\overline K_2 + T_r$ has $\chi=3$ and $q_c=5/2 < \chi$, and 
$\overline K_2 + (Ch)_{3,r}$ has $\chi=4$ and $q_c=7/2 < \chi$.  We shall
discuss the situation for homeomorphic expansions of graphs in these families
below.

\vspace{10mm}

\section{Families of the Form $(K_{\lowercase{p}})_{ \{ \lowercase{b} \}} + 
G_{\lowercase{r}}$}

Another way to construct families with noncompact $W$ boundaries ${\cal B}$ is
to remove bonds not just from one vertex of the $K_p$ subgraph but also from 
one or more other vertices of this subgraph that are not adjacent to the first
vertex, i.e., are not connected to this first vertex by any bonds.  We
symbolize this family by 
\beq
(K_p)_{ \{b \}} + G_r
\label{kpgmultipleb}
\eeq
where the subscript $\{b\}$ refers to the removal of multiple bonds from
non-adjacent vertices of the $K_p$ subgraph.  
For this family, the maximum number of bonds 
that can be removed from the $K_p$ subgraph is the number of bonds in
this subgraph, namely, 
\beq
e(K_p) = {p \choose 2} = \frac{p(p-1)}{2}
\label{ekp}
\eeq
In general, the chromatic polynomial for such families is not quite
as easily derived as that for the type of families where all bonds removed are
from one vertex, where we obtained the simple formula (\ref{pkpcutb})
\cite{wa}.  

A particularly simple family of graphs
is obtained when one removes all of the 
$p \choose 2$ bonds in the $K_p$ subgraph, thereby yielding the subgraph
$\overline K_p$ and the resultant family
\beq
\overline K_p + G_r
\label{epg}
\eeq
Clearly, for $p=2$, 
\beq
 (K_2)_{b=1} + G_r = \overline K_2 + G_r
\label{e2grrel}
\eeq
For the general family $\overline K_p + G_r$, one can use the 
addition-contraction \cite{thm} 
theorem to reexpress $P(\overline K_p + G_r,q)$ as the linear combination
\beqs
P(\overline K_p + G_r,q) & = & \sum_{j=1}^p {\cal S}_p^{(j)}P(K_j+G_r,q) \cr
                  & = & \sum_{j=1}^p {\cal S}_p^{(j)} q^{(j)}P(G_r,q-j)
\label{pepk}
\eeqs
where the coefficients ${\cal S}_p^{(j)}$ are the Stirling numbers of the 
second kind, defined by the equation (cf. eq. (\ref{ff}))
\beq
q^p = \sum_{j=1}^p {\cal S}_p^{(j)} q^{(j)}
\label{qpff}
\eeq
A closed-form expression is \cite{combin} 
\beq
{\cal S}_p^{(j)} = \frac{1}{j!} \sum_{k=0}^j (-1)^{j-k}{j \choose k} k^p
\label{stirlings2}
\eeq
Given that $P(G,q)$ 
has the form (\ref{pgsum}), it follows, by the same argument as
was used for the $(K_p)_b + G_r$ families with removal of $b$ bonds connecting
to a given vertex, that ${\cal B}$ is noncompact in the $q$ plane, 
passing through $z=1/q=0$.  

It is worthwhile to consider some special cases for $G_r$.  For 
$G_r= \overline K_r$, we calculate 
\beq
P(\overline K_p + \overline K_r,q) = 
\sum_{j=1}^p {\cal S}_p^{(j)}q^{(j)}(q-j)^r
\label{peper}
\eeq
Hence, 
in the limit $r \to \infty$, ${\cal B}$ is determined by the degeneracy
equation
\beq
|q-1|=|q-p| \ , \quad i.e., \quad |1-z|=|1-pz|
\label{degeqeper}
\eeq
Clearly, $z=0$ is a solution, and ${\cal B}$ is, in the $q$ plane, the vertical
line with
\beq
Re(q) = q_c = \frac{p+1}{2}
\label{qceper}
\eeq
i.e., in the $z$ plane the circle 
\beq
|z- \frac{z_c}{2}| = \frac{z_c}{2} \ , \quad z_c = \frac{2}{p+1}
\label{bzeper}
\eeq
Let us compare the values of $q_c$ for the respective $r \to \infty$ limits of
$\overline K_p + \overline K_r$ and $(K_p)_b + \overline K_r$ (where it is
understood in the latter case that $b$ bonds are removed from a
vertex in the $K_p$ subgraph). 
For $p=2$, the value of $q_c$ in eq. (\ref{qceper}) is equal to that
for $G_r = \overline K_r$ in eq. (\ref{qcthreecases}), 
as follows from eq. (\ref{e2grrel}).  For $p \ge 3$, the
value of $q_c$ in eq. (\ref{qceper}) is smaller than that for 
$G_r = \overline K_r$ in eq. (\ref{qcthreecases}).  This is a consequence of
the fact that the graph $(K_p)_b + \overline K_r$ can be obtained from 
$\overline K_p + \overline K_r$ by addition of bonds and hence, by the 
theorem in section VII of Ref. \cite{wn}, 
the coloring of the former graph with $q$ colors is
more constrained than the coloring of the latter graph; the same theorem
implies that for $q \in {\mathbb Z}_+$, 
\beq
W(\lim_{r \to \infty}[(K_p)_b + \overline K_r],q) \le 
W(\lim_{r \to \infty}[\overline K_p + \overline K_r],q) 
\label{wkpkrrel}
\eeq
In region $R_1$ to the right of the vertical line (\ref{qceper}) in the $q$
plane, 
\beq
W(\lim_{r \to \infty}(\overline K_p + \overline K_r),q) = q-1 
\quad {\rm for} \quad q \in R_1
\label{weperregion1}
\eeq
while in the complement, $R_2$, 
\beq
|W(\lim_{r \to \infty}(\overline K_p + \overline K_r),q)| = |q-p| 
\quad {\rm for} \quad q \in R_2
\label{weperregion2}
\eeq

A second case is $G_r=T_r$, i.e., the family $\overline K_p + T_r$.  We find 
\beq
P(\overline K_p + T_r,q) = \sum_{j=1}^p {\cal S}_p^{(j)}q^{(j)}
(q-j)(q-j-1)^{r-1}
\label{peptr}
\eeq
Hence, in the limit $r \to \infty$, ${\cal B}$ is determined by the degeneracy
equation
\beq
|q-2|=|q-p-1| \ , \quad i.e., \quad |1-2z|=|1-(p+1)z|
\label{degeqeptr}
\eeq
In the $q$ plane the locus of solutions to this equation is the vertical line
with
\beq
Re(q) = q_c=\frac{p+3}{2}
\label{qceptr}
\eeq
which again is a circle in the $z$ plane, 
\beq
|z- \frac{z_c}{2}| = \frac{z_c}{2} \ , \quad z_c = \frac{2}{p+3}
\label{bzeptr}
\eeq
For $p=2$, the value of $q_c$ in eq. (\ref{qceptr}) is equal to that
for $G_r = T_r$ in eq. (\ref{qcthreecases}),
as follows from eq. (\ref{e2grrel}).  For $p \ge 3$, $q_c$ 
is larger for $\lim_{r \to \infty}[(K_p)_b + T_r]$ with $b$ bonds removed from
one vertex in $K_p$ than for $\lim_{r \to \infty}[\overline K_p + T_r]$, for
the reason given above.  For region $R_1$ to the right of the vertical line 
(\ref{qceptr}) in the $q$ plane, 
\beq
W(\lim_{r \to \infty}[\overline K_p + T_r],q) = q-2 
\quad {\rm for} \quad q \in R_1
\label{weptrregion1}
\eeq
while in the complement, $R_2$, 
\beq
|W(\lim_{r \to \infty}[\overline K_p + T_r],q)| = |q-p-1|
\quad {\rm for} \quad q \in R_2
\label{weptrregion2}
\eeq

\vspace{10mm}

\section{Construction of Families of Graphs with Noncompact ${\cal B}$ via
Homeomorphic Expansion}

A major result of the present paper is the construction of families of graphs
with noncompact $W$ boundaries by means of homeomorphic expansion ($HE$) (also
called inflation) of a beginning set of families of graphs.  We recall the 
definition that two graphs $G$ and $H$ are homeomorphic to each other if $H$, 
say, can be obtained from $G$ by successive insertions of degree-2 vertices 
on bonds of $G$ \cite{biggsbook}.  
Each such insertion subdivides an existing edge of $G$ into two,
connected by the inserted degree-2 vertex.  This process is called 
homeomorphic expansion and its inverse is called 
homeomorphic reduction, i.e. the successive removal of vertices of degree 2
from a graph $H$.  Clearly, homeomorphic expansion of a graph always yields
another graph.  The inverse is not necessarily true; i.e., homeomorphic
reduction of a graph can produce a multigraph or pseudograph instead of a 
(proper) graph \cite{graphdef}.  Here, a multigraph is a finite set of 
vertices and bonds that, like a graph, has no bonds that loop around from a
given vertex back to itself but, in contrast to a (proper) graph, 
may have more than one bond connecting two vertices.  A pseudograph is a finite
set of vertices and bonds that may have multiple bonds connecting two
vertices and may also have looping bonds.  For example, consider homeomorphic
reduction of a circuit graph $C_r$: removing one of the vertices (all of 
which have degree 2), one goes from $C_r$ to $C_{r-1}$, and so forth, until one
gets to $C_3$.  During this sequence of homeomorphic reductions, one remains
within the category of graphs.  However, the next homeomorphic reduction takes
$C_3$ to $C_2$, which is a multigraph, not a proper graph.  Removing one of the
two vertices in $C_2$ produces a pseudograph consisting of a single vertex and
a bond that goes out and loops back to this vertex.  Thus homeomorphism is an
equivalence relation on pseudographs.  This complication will not 
be relevant here because we shall only use homeomorphic expansions, not
reductions, of graphs, and the homeomorphic expansion of a proper graph always
yields another proper graph.  For our subsequent discussion, we shall
denote the homeomorphic expansion involving the insertion of $k-2$ additional
vertices (where $k \ge 3$)
on a specific bond $b$ of a graph $G$ as $HEG_{k-2;b}(G)$.
Most of our studies will be of graphs in which the homeomorphic expansion of
$G$ is performed systematically on each bond of $G$; in these cases, we shall
denote the resultant graph as $HEG_{k-2}(G)$.

The homeomorphic expansion of a family of graphs with a compact (empty or 
nontrivial) locus ${\cal B}$ can produce a family of graphs with either a 
compact or noncompact  ${\cal B}$; this depends on the nature of the original
family and of the homeomorphic expansion. 
For example, if one starts with the tree
graph $T_r$ and inserts $k-2$ degree-2 vertice on each of the bonds, one 
obtains another tree graph, $T_{r'}$, where $r'=r+(r-1)(k-2)$.  In the limit
$r \to \infty$, both $T_r$ and its homeomorphic expansion, $T_{r'}$, 
have a trivial ${\cal B}=\emptyset$. 
If one starts with the circuit graph $C_r$ and inserts $k-2$ degree-2 vertices
on each bond of this graph, one obtains another circuit graph, $C_{r'}$, where 
$r'=r+r(k-2)$.  In the limit $r \to \infty$, both $C_r$ and its homeomorphic
expansion $C_{r'}$ have a compact ${\cal B}$ given by the unit circle centered
at $q=1$, $|q-1|=1$.  We next proceed to discuss the cases of main interest
here, where the homeomorphic expansion (i) leads from a family with a compact
${\cal B}$ to one with a noncompact ${\cal B}$ or (ii) takes a family that 
already has a noncompact ${\cal B}$ to another that again has a noncompact
${\cal B}$.  We comment on the effect of a homeomorphic expansion of a graph 
$G$ on its girth $\gamma(G)$, defined as the length of (= number of vertices 
on) a minimal-length closed circuit in this graph.  Clearly, a homeomorphic 
expansion applied to one or more bonds of a graph increases the girth of the 
graph if and only if these bonds lie on the minimal-length circuits of the
original graph. 
We shall consider four main types of homeomorphic expansions, 
as well as combinations thereof: 

\vspace{10mm}

\begin{flushleft}

1. \ $HEK0$ 

\vspace{6mm} 

Start with the family $K_p + G_r$ (where no bonds have been removed from any
vertex of the $K_p$ subgraph).  If in the limit $r \to \infty$, ${\cal B}$ for
the $G_r$ family itself is compact (bounded) in the $q$ plane, then the same
holds for the family $K_p + G_r$.  Now insert degree-2 vertices on one or more 
bonds of the $K_p$. We shall denote this generically as an $HEK0$ homeomorphic
expansion, meaning that the homeomorphic expansion 
acts on the $K_p$ subgraph and that originally
there were $b=0$ bonds removed from this $K_p$.  Specifically, we
denote the graph obtained by successive insertion of $k-2$ degree-2 vertices on
a single bond $b_j$ 	of the $K_p$ as 
\beq
HEK_{k-2;b_j}(K_p + G_r) 
\label{hekb0}
\eeq
where $k \ge 3$, 
and so forth for similar homeomorphic insertions on other bonds of the $K_p$. 
We will show below that this homeomorphic expansion leads from the family 
$K_p + G_r$ with compact (trivial or nontrivial) ${\cal B}$ to the family
(\ref{hekb0}) with a locus ${\cal B}$ that is noncompact in the $q$ plane. 
The labelling convention is chosen so that as one moves along the expanded set
of bonds linking a vertex of the original $K_p$ to what was originally an
adjacent vertex, one traverses a total of $k$ vertices, including the original
pair, i.e., $k-2$ inserted vertices. 

\vspace{8mm}

Next, we consider cases (nos. 2-5 below) 
where we begin with a family of graphs that already has
a locus ${\cal B}$ that is unbounded in the $q$ plane.  For these the
homeomorphic expansion produces another family of graphs again with an
unbounded ${\cal B}$: 

\vspace{8mm}

2. \ $HEK$

\vspace{6mm}

Start with the family $(K_p)_b + G_r$ where $b$ bonds have been removed from
one vertex of the $K_p$ subgraph and insert degree-2 vertices on one or more 
bonds of the $K_p$. We shall denote this generically as an $HEK$ homeomorphic
expansion, meaning that the homeomorphic expansion 
acts on the $K_p$ subgraph.  Specifically, we
denote the graph obtained by successive insertion of $k-2$ degree-2 vertices on
a single bond of the $K_p$ as 
\beq
HEK_{k-2;b_j}[(K_p)_b + G_r]
\label{hekb}
\eeq 
where again $k \ge 3$, 
and so forth for similar insertions on other bonds of the $K_p$ subgraph. 

\vspace{8mm}

3. \ $HEG$

\vspace{6mm}

Start with either $(K_p)_b + G_r$ or $(K_p)_{\{b\}} + G_r$ 
and add vertices to bonds in the $G_r$ subgraph. 
We label this type of homeomorphic expansion generically as $HEG$. 
Analogously to the previous category, we denote the respective graphs 
obtained by successive insertion of $k-2$ degree-2 vertices on
a single bond of the $G_r$ subgraph as 
\beq
HEG_{k-2;b_j}[(K_p)_b + G_r] \ , \quad 
HEG_{k-2;b_j}[(K_p)_{\{b\} } + G_r]
\label{hegb}
\eeq
and the graphs obtained by successive insertions of $k-2$ degree-2 vertices on
all of the bonds of $G_r$ as 
\beq
HEG_{k-2}[(K_p)_b + G_r] \ , \quad 
HEG_{k-2}[(K_p)_{\{ b \} } + G_r] 
\label{heg}
\eeq

\vspace{8mm}

4. \ $HEC$

\vspace{6mm}

Start with either $(K_p)_b + G_r$ or $(K_p)_{\{b\}} + G_r$ 
and add vertices to the bonds connecting vertices in
the $K_p$ subgraph to vertices in the $G_r$ subgraph. We label this type of
homeomorphic expansion generically as $HEC$, where the ``C'' refers to the fact
that the homeomorphic expansion is performed on the above-mentioned connecting
bonds.  We denote the respective graphs 
obtained by successive insertion of $k-2$ degree-2 vertices on
a single bond $b_{ij}$ connecting a vertex $v_i \in K_p$ to a vertex 
$v_j \in G_r$ as 
\beq
HEC_{k-2; b_{ij}}[(K_p)_b + G_r] \ , \quad 
HEC_{k-2; b_{ij}}[(K_p)_{\{ b \} } + G_r]
\label{hecij}
\eeq
and the graphs obtained by successive insertions of $k-2$ degree-2 vertices on
all of the bonds connecting vertices of $K_p$ to vertices of $G_r$ as
\beq
HEC_{k-2}[(K_p)_b + G_r] \ , \quad 
HEC_{k-2}[(K_p)_{\{ b \} } + G_r ]
\label{hec}
\eeq

\vspace{8mm}

5. \ Combinations

\vspace{6mm}

Clearly, one can combine several types of homeomorphic expansion.  For example,
starting with $(K_p)_b + G_r$, one can add vertices both to bonds in the 
subgraph $K_p$, to bonds that connect $K_p$ to $G_r$, and to bonds in the $G_r$
subgraph. 

\end{flushleft} 

In the present paper we shall concentrate on homeomorphic expansions of types
(1)-(3).  Our results for homeomorphic expansions of type (4) 
involve somewhat more complicated boundaries ${\cal B}$ than those discussed
here and will be dealt with in a separate paper.  Composite homeomorphic 
expansions of type (5) can be studied by similar means.  We now proceed to 
consider the various homeomorphic classes in more detail. 

Given that the number of vertices in the original graph is a linear function of
the two (positive integer) parameters $p$ and $r$, the number $n$ of vertices 
of the homeomorphic expansion is a linear function of $p$, $r$, and $k$.  There
are therefore three basic ways of producing the limit $n \to \infty$, namely 
($L$ denotes limit) 
\beq
L_p: \ p \to \infty \quad {\rm with} \quad r \quad {\rm and} \quad k \quad 
{\rm fixed}
\label{pinf}
\eeq
\beq
L_r: \ r \to \infty \quad {\rm with} \quad p \quad {\rm and} \quad k \quad
{\rm fixed}
\label{rinf}
\eeq
\beq
L_k: \ k \to \infty \quad {\rm with} \quad p \quad {\rm and} \quad r \quad
{\rm fixed}
\label{kinf}
\eeq
We have explained above (after eq. (\ref{mageq})) why the limit $L_p$ is not 
very interesting.  From the viewpoint of the present work on boundaries 
${\cal B}$ that are noncompact in the $q$ plane, the limit $L_k$ is also not 
of primary interest, since it generically yields a compact boundary 
${\cal B}$, as we shall illustrate below.  Hence we shall concentrate on the 
limit $L_r$ here.

\section{$HEK0$ and $HEK$ Homeomorphic Classes} 

Let us consider first the homeomorphic expansions of type (1), namely, 
$HEK_{k-2,b_j}(K_p+G_r)$.  
By use of the deletion-contraction theorem and eq. (\ref{pkpg}), we 
find that, for arbitrary $G_r$, 
\beqs
P \Bigl ( HEK_{k-2,b_j}(K_p+G_r),q \Bigr ) & = & D_k P(K_p + G_r,q) + 
[D_k + (-1)^{k-1}]P(K_{p-1} + G_r,q) \cr
& = & D_k q^{(p)} P(G_r,q-p) + [D_k + (-1)^{k-1}] q^{(p-1)}P(G_r,q-(p-1))
\label{phkkpg}
\eeqs
It follows by the same argument as that given with eq. (\ref{pkpcutb}) that as
$r \to \infty$, this family $HEK_{k-2 \ge 1,b_j}(K_p+G_r)$ has a $W$ boundary 
${\cal B}$ that is noncompact in the $q$ plane. 
This is true independent of whether the family $G_r$ has
this property.  Indeed, as is evident from a comparison of eq. (\ref{phkkpg})
with eq. (\ref{pkpcutb}), for a given $G_r$, 
\beq
{\cal B}\Bigl ( \lim_{r \to \infty} HEK_{k-2,b_j}(K_p+G_r) \Bigr ) = 
{\cal B}\Bigl ( \lim_{r \to \infty} [(K_p)_b + G_r] \Bigr )
\label{bsame}
\eeq
That is, if we start with the family $K_p + G_r$, remove $b$ bonds from a 
given vertex in the $K_p$ subgraph, and take $r \to \infty$, the resultant 
boundary ${\cal B}$ is the same as if we had instead homeomorphically added 
some number $k-2 \ge 1$ degree-2 vertices to a bond in the $K_p$ subgraph.  One
can, of course, continue this process with homeomorphic expansions of other
bonds of the $K_p$.  Note that if one uses the limit $L_k$ in eq. 
(\ref{kinf}) to get $n \to \infty$ for this class of homeomorphic expansions,
then, for arbitrary 
(finite) $G_r$, the resultant boundary is the compact locus comprised by the
unit circle $|a|=1$ (where $a$ was defined in eq. (\ref{a}), i.e., 
\beq
{\cal B}_{L_k} = q \quad {\rm such \ \ that} \quad |q-1|=1
\label{blk}
\eeq

\vspace{6mm}

We next consider the homeomorphic expansions of type (2), namely, 
$HEK_{k-2;b_j}[(K_p)_b + G_r]$.  Here, the boundary ${\cal B}$ for the 
$r \to \infty$ of the beginning family $(K_p)_b + G_r$ is
already noncompact in the $q$ plane, and the families generated by the
homeomorphic expansion maintain this property.  As an illustration, consider
the family $(K_3)_1 + G_r$ and homeomorphically expand one of the two remaining
bonds, denoted $b_j$, in the $(K_3)_1 = T_3$ subgraph. 
We find for the resultant chromatic
polynomial the result
\beqs
P(HEK_{k-2;b_j}[(K_3)_1 + G_r],q) & = & D_k q^{(3)}P(G_r,q-3) \cr\cr
 & & + [2D_k + (-1)^{k-1}]q^{(2)}P(G_r,q-2) 
\label{hekexample}
\eeqs 
Again, using the same reasoning that we employed before with eq. 
(\ref{pkpcutb}), we deduce that as $r \to \infty$, this family 
$HEK_{k-2;b_j}[(K_3)_1 + G_r]$ has a noncompact locus ${\cal B}$ in the $q$
plane.  As was the case for the $HEK0$ class, for the limit $L_k$, the 
resultant boundary is the unit circle $|q-1|=1$ given in eq. (\ref{blk}).

\section{The Family $T_{\lowercase{k,r}} = 
HEG_{\lowercase{k}-2}(\overline K_2 +T_{\lowercase{r}})$} 

We next proceed to analyze in detail several homeomorphic classes of graph
families of the form $HEG$.  One 
interesting family of graphs with noncompact ${\cal B}(q)$ is obtained by 
homeomorphic expansion starting with the family $(K_p)_b + T_r$ for 
$r \ge 2$ and adding vertices on each of the bonds in the $T_r$ subgraph.  
We let $k$ be the number of vertices on $T_r$ between each pair of vertices
that connect with the $K_p$, including this pair (which were originally
adjacent on $T_r$ before the homeomorphic expansion).  
We denote this family as 
\beq
T_{p,b,k,r} = HEG_{k-2}[(K_p)_b + T_r]
\label{tpbkr}
\eeq
where as above, $r \ge 2$ and $1 \le b \le p-1$.  
The number of vertices is given by
\beq
v(T_{p,b,k,r}) = (r-1)(k-1)+p+1
\label{etpbkr}
\eeq
For the lowest value $k=2$, 
\beq
T_{p,b,k=2,r} = (K_p)_b + T_r
\label{tpbkeq2r}
\eeq
which we studied previously \cite{wa}. 
We shall thus concentrate on the cases $k \ge 3$ here. 
It suffices for our present purposes to consider the 
simplest nontrivial case $p=2$ and hence $b=1$, for which 
\beq
(K_2)_{b=1} + T_r = \overline K_2 + T_r
\label{k2bartr}
\eeq
An illustration of a graph of this type is given in Fig. \ref{vhegraph1}(a). 
For brevity, we define 
\beq
T_{k,r} = T_{p=2,b=1,k,r} = HEG_{k-2}(\overline K_2 + T_r) 
\label{tkrdef}
\eeq
We observe that the chromatic number is 
\beqs
\chi(T_{k,r}) & = & 2 \quad {\rm for} \quad k \ \ {\rm odd} \cr
                  & = & 3 \quad {\rm for} \quad k \ \ {\rm even} 
\label{chitkr}
\eeqs
The girth, i.e., the length of a minimal-length circuit on this graph, is 
\beq
\gamma(T_{k,r}) = k+1
\label{girthtkr}
\eeq
By the deletion-contraction theorem, we find the recursion relation 
\beq
P(T_{k,r},q) = \Bigl [ D_3D_k + (-1)^{k-1} \Bigr ] P(T_{k,r-1},q) + 
q(q-1)D_k (D_{k+1})^{r-2}
\label{tkrrecursion}
\eeq
Solving this, we obtain the chromatic polynomial
\beq
P(T_{k,r},q) = A_k \Bigl [  D_3D_k + (-1)^{k-1} \Bigr ]^{r-2}
+ q(q-1)(D_{k+1})^{r-1}
\label{ptkr}
\eeq
where
\beq
A_k = (q-3)P(C_{k+1},q) + (q-1)P(C_{k-1},q) 
\label{ak}
\eeq
$P(T_{k,r},q)$ has the general factors $q(q-1)$ for $k$ odd and 
$q(q-1)(q-2)$ for $k$ even.  As is evident from eq. (\ref{ptkr}), 
$P(T_{k,r},q)$ has the form of eq. (\ref{pgsum}) with 
\beq
a_1 = D_{k+1}   
\label{a1tkr}
\eeq
and 
\beq
a_2 = D_3D_k + (-1)^{k-1}
\label{a2tkr}
\eeq
(recall that $D_3=q-2$).  In the limit $r \to \infty$ (with $k$ fixed 
\cite{lknote}) 
the locus ${\cal B}$ is determined by the degeneracy of magnitudes 
\beq
|a_1|=|a_2|
\label{degeneqtkr}
\eeq
This equation can be simply expressed in terms of the variable $a=q-1$ which
was defined above in eq. (\ref{a}) (and should not be confused with the
variables $a_1(q)$ and $a_2(q)$).  To do this, we multiply both sides of eq. 
(\ref{degeneqtkr}) by $|q(q-1)|=|(a+1)a|$ (the spurious solutions at $q=0$ 
and $q=1$ thereby introduced are discarded) and use eq. (\ref{pck}) to get 
\beq
|a(a^k - a^{k-1} + 2(-1)^{k+1})| = |a(a^k + (-1)^{k+1})|
\label{aeqtkr}
\eeq
i.e., after dividing by $|a|^{k+1}$,
\beq
|1-y+2(-1)^{k+1}y^k| = |1+(-1)^{k+1}y^k|
\label{yeqtkr}
\eeq
Since $y=0$ is a solution of this equation, ${\cal B}$ is noncompact in the $q$
plane, passing through $y=z=0$.  In polar coordinates, with $y=\rho
e^{i\beta}$, eq. (\ref{yeqtkr}) yields 
\beq
\rho\Bigl [3\rho^{2k-1} + \rho - 2\cos \beta + 2(-1)^{k-1}\rho^{k-1} 
\biggl \{ \cos(k\beta) - 2\rho \cos((k-1)\beta) \biggr \} \Bigr ] = 0
\label{yeqtkrpolar}
\eeq
As $\rho \to 0$, it follows that $\cos\beta=0$, i.e. $\beta = \pm \pi/2$, so
that ${\cal B}$ approaches $y=z=0$ vertically.

To calculate the point at which the boundary ${\cal B}$ crosses the real $q$,
or equivalently, $y$ or $z$ axes, we set $\beta=0$ in
eq. (\ref{yeqtkrpolar}); for $\rho \ne 0$ this gives
\beq
3\rho^{2k-1} + \rho - 2 + 2(-1)^{k-1}\rho^{k-1}(1-2\rho)=0
\label{yeqtkrpolarpositivey}
\eeq
(Note that $Re(q) > 0 \Longleftrightarrow 
Re(z) > 0$ and $Re(q) > 1 \Longleftrightarrow Re(y) > 0$ so that setting
$\beta=0$, i.e. making $y$ real and positive, implies that $q$ is real and $q >
1$; we comment later on the interval $0 < q < 1$.)  Since $\rho=|y|$ is real 
and positive, we are only interested in roots of eq. 
(\ref{yeqtkrpolarpositivey}) 
that have this property.  For both even and odd $k$, there is only one such
root, which is thus $y_c$ (equivalently, $z_c$), the minimum nonzero value of
$y$ ($z$) respectively, at which the boundary ${\cal B}$ crosses the
positive real axis in the $y$ ($z$) plane, corresponding to the maximal finite
real point $q_c$ at which the pre-image of this boundary ${\cal B}$ crosses the
positive real axis in the $q$ plane.  For $k$ odd, this single root is 
$\rho=y=1$ (the expression on the left-hand side contains a factor 
$(\rho-1)$) so that $z_c=1/2$, and 
\beq
q_c = 2 = \chi \quad {\rm for} \quad k \quad {\rm odd}
\label{qce2tkr_kodd}
\eeq
For $k$ even, the single such root of eq. (\ref{yeqtkrpolarpositivey}) 
has a value that increases monotonically from $\rho=y_c=2/3$ (equiv. 
$z_c=2/5$) for $k=2$ toward the limit $\rho=y_c=1$ (equiv. 
$z_c=1/2$) as $k \to \infty$ through even integers.  Correspondingly, for $k$
even, $q_c$ decreases monotonically from 5/2 for $k=2$ toward 2 in the same
limit; some values are 
$q_c=2.2564, \ 2.1736, \ 2.1315$ for $k=4, \ 6, \ 8$).  Hence, 
\beq
2 < q_c < \frac{5}{2} < \chi = 3 \quad {\rm for} \quad k \quad {\rm even}
\label{qce2tkr_keven}
\eeq

To show that ${\cal B}$ does not cross the positive real $q$ axis in the
interval $0 \le q < 1$, we take $\beta=\pi$, i.e., $y$ real and negative, in
eq. (\ref{yeqtkrpolar}), which, for $\rho \ne 0$, yields the equation
\beq
3\rho^{2k-1}+\rho+2-2\rho^{k-1}-4\rho^k=0
\label{betapieq}
\eeq
This range $-\infty < y \le 0$
includes both the interval $0 \le q < 1$ and the interval $-\infty < q < 0$.
We find that in this range of $y$, eq. (\ref{betapieq}) has no 
real positive roots for $\rho$ except for $\rho=1$, i.e., $y=-1$ or
equivalently, $q=0$, but this is just a spurious solution introduced when we
multiplied both sides of eq. (\ref{degeneqtkr}) by $|q(q-1)|$ to express it as
eq. (\ref{aeqtkr}).  Hence, ${\cal B}$ crosses
the positive $z$ or equivalently $q$ axis only once for this family of graphs. 

We proceed to discuss the boundary ${\cal B}$ for specific values of $k$ 
further.  For $k=2$ \cite{wa} 
${\cal B}$ is the vertical line with $Re(q)=5/2$ in the $q$
plane, or equivalently, the circles $|z-z_c/2|=z_c/2$, $|y-y_c/2|=y_c/2$ in the
$z$ and $y$ planes, where $z_c=2/5$, $y_c=2/3$ as given above. 

\pagebreak

\begin{figure}
\vspace{-4cm}
\centering
\leavevmode
\epsfxsize=3.0in
\begin{center}
\leavevmode
\epsffile{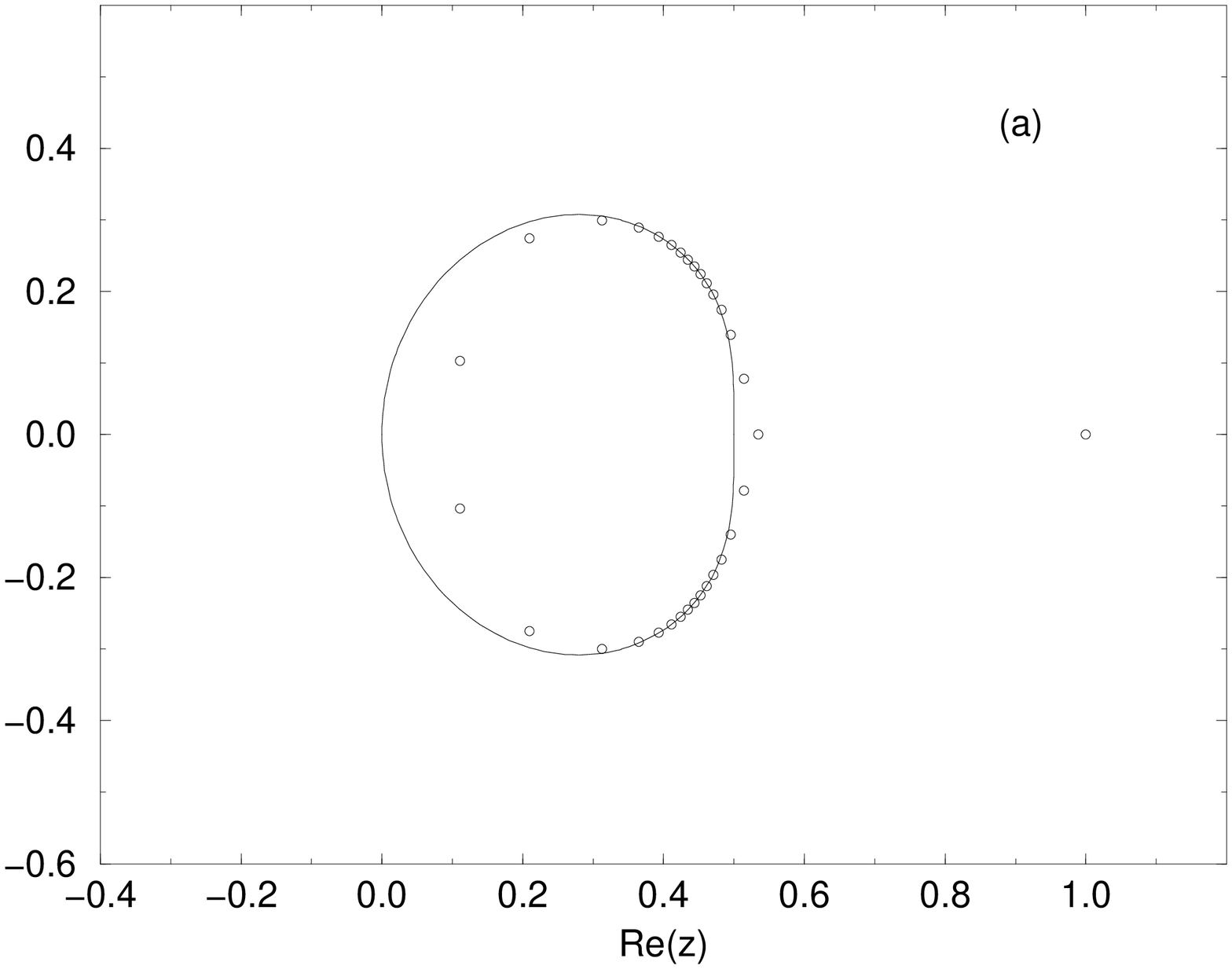}
\end{center}
\vspace{-2cm}
\begin{center}
\leavevmode
\epsfxsize=3.0in
\epsffile{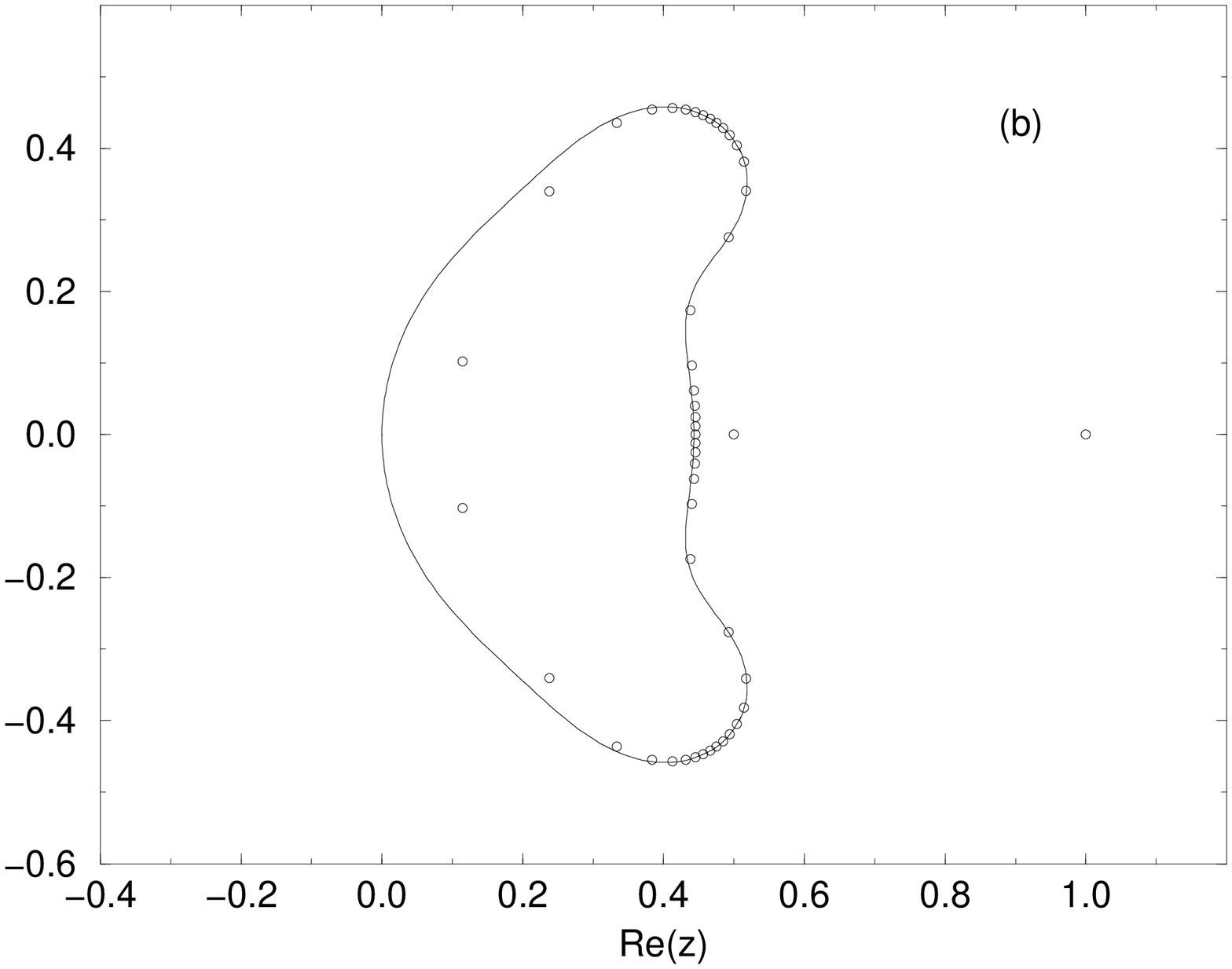}
\end{center}
\vspace{-2cm}
\caption{\footnotesize{Boundary ${\cal B}$ in the $z=1/q$ plane for the $r \to
\infty$ limit of the family of graphs $HEG_{k-2}(\overline K_2 + T_r)$ 
with $k=$ (a) 3 (b) 4. Chromatic zeros for $r=16$ are shown for 
comparison.}}
\label{boundarye2hvktr1}
\end{figure}

\pagebreak

\begin{figure}
\vspace{-4cm}
\centering
\leavevmode
\epsfxsize=3.0in
\begin{center}
\leavevmode
\epsffile{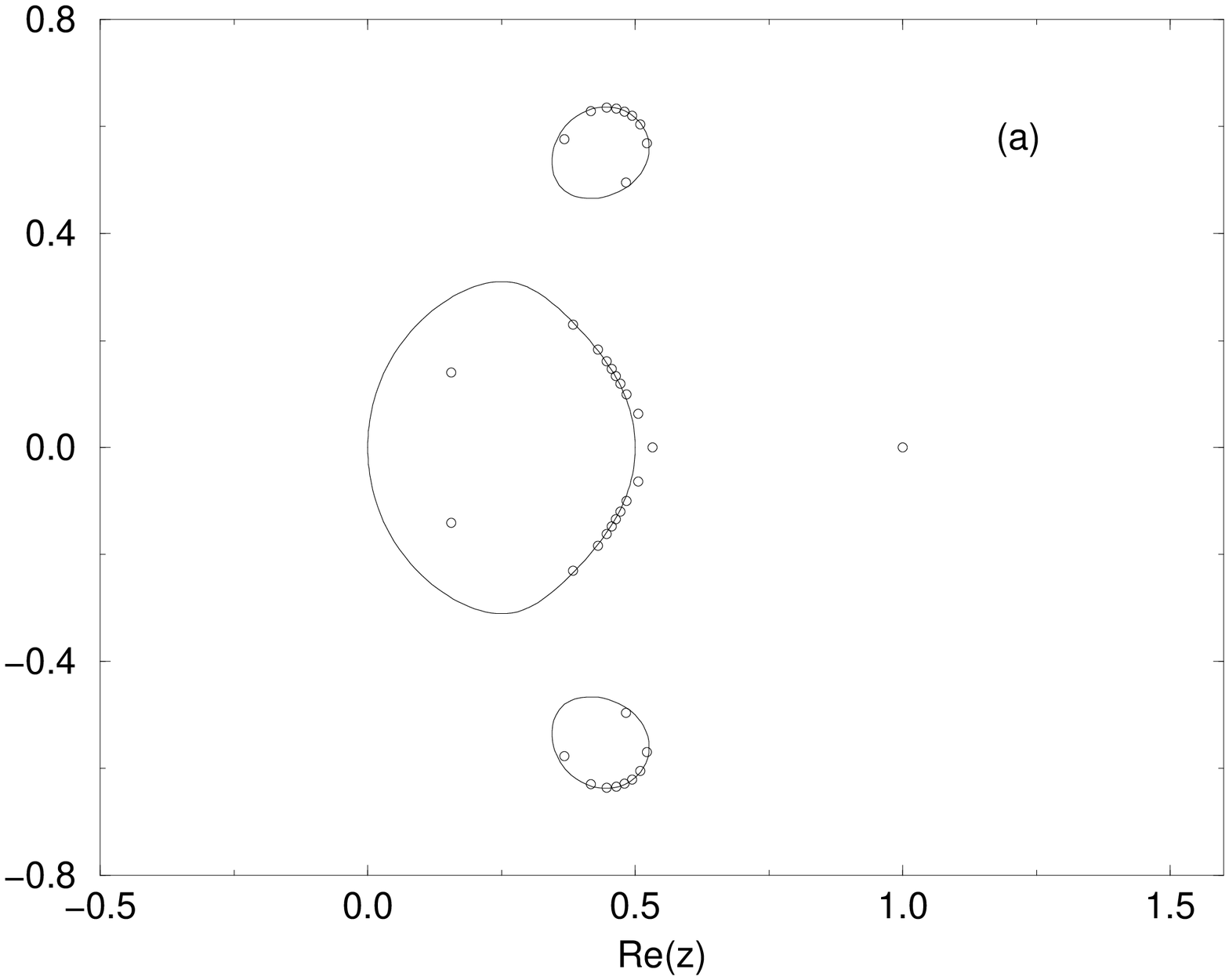}
\end{center}
\vspace{-2cm}
\begin{center}
\leavevmode
\epsfxsize=3.0in
\epsffile{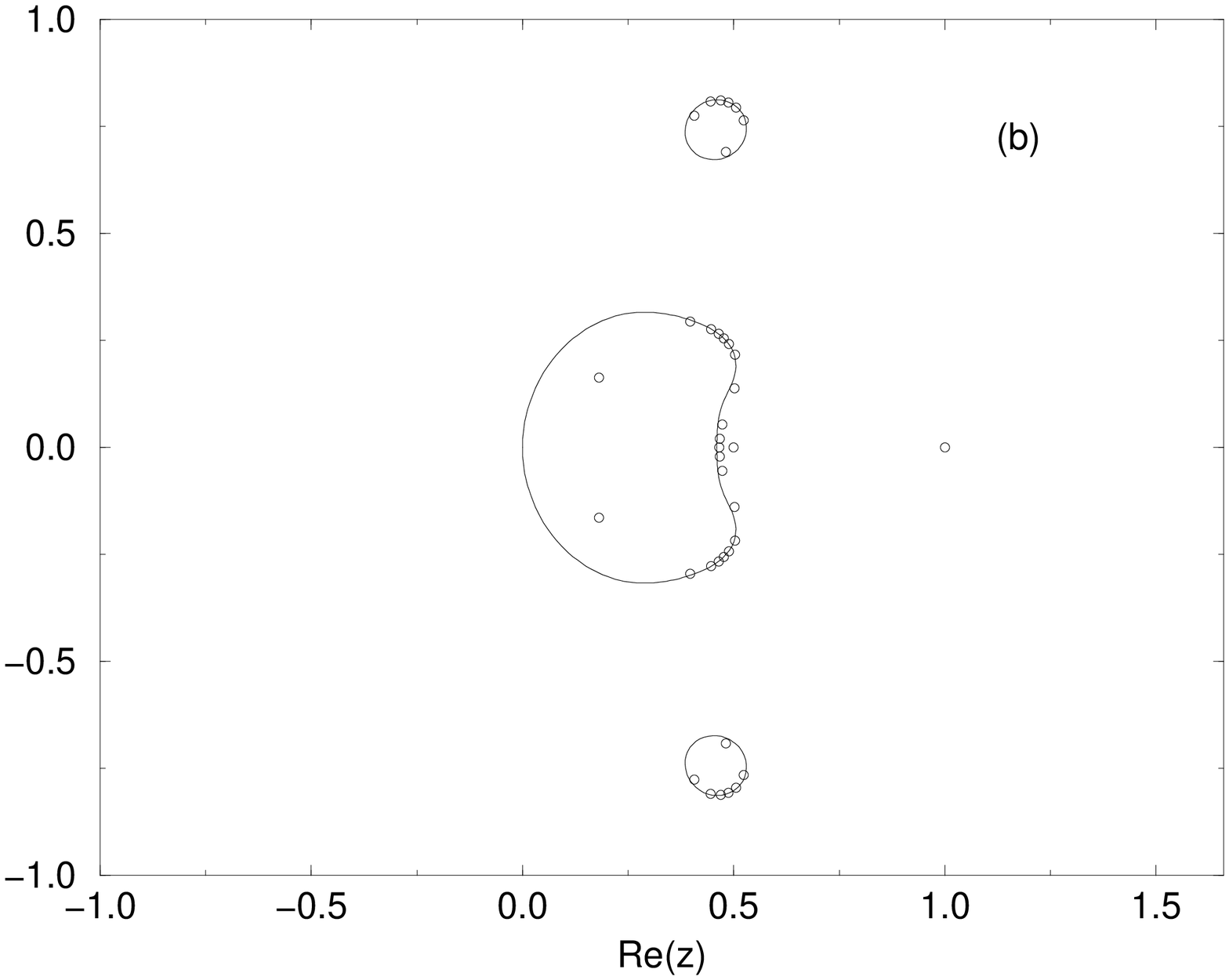}
\end{center}
\vspace{-2cm}
\caption{\footnotesize{As in Fig. \ref{boundarye2hvktr1} for $k=$ (a) 5 (b) 6.
Chromatic zeros for $r=$ (a) 10 (b) 8 are shown for comparison.}}
\label{boundarye2hvktr2}
\end{figure}

The boundary ${\cal B}$ is shown in the $z$ plane for $3 \le k \le 6$ in 
Figs. \ref{boundarye2hvktr1} and \ref{boundarye2hvktr2} \cite{thesis}.  
In region $R_1$
containing the portion of the positive $z$ axis analytically connected to the
point $z=0$, 
\beq
W([\lim_{r \to \infty} T_{k,r}],q) = (a_1)^{1/(k-1)} = (D_{k+1})^{1/(k-1)}
 \quad {\rm for} \quad q \in R_1
\label{wtkr_r1}
\eeq
In region $R_2$ occupying the rest of the $z$ plane for $2 \le k \le 4$, and
the rest of the $z$ plane except for the interiors of the other closed loops
for $k \ge 5$, 
\beq
|W([\lim_{r \to \infty} T_{k,r}],q)| = |a_2|^{1/(k-1)} = |D_3D_k +
(-1)^{k-1}|^{1/(k-1)} \quad {\rm for} \quad q \in R_2
\label{wtkr_r2}
\eeq
For odd $k$, as $q$ decreases toward 2 one has, using eqs. (\ref{wtkr_r1}) 
and (\ref{dkoddq2}), 
\beq
\lim_{q \searrow 2}W([\lim_{r \to \infty} T_{k,r}],q) = 1
\label{tkrwq2region1}
\eeq
and, as $q$ increases toward 2, one has, using eq. (\ref{wtkr_r2}), 
\beq
\lim_{q \nearrow 2}||W([\lim_{r \to \infty} T_{k,r}],q)| =1
\label{tkrwq2region2}
\eeq
The equality of the right-hand sides of eqs. (\ref{tkrwq2region1}) and 
(\ref{tkrwq2region2}) follows from the fact that $q=2$ is on the boundary
${\cal B}$ for odd $k$. 
In each of the additional regions enclosed by closed curves, 
\beq
|W([\lim_{r \to \infty} T_{k,r}],q)| = |a_1|^{1/(k-1)} = |D_{k+1}|^{1/(k-1)}
\label{wtkr_r3}
\eeq
For comparison, we show chromatic zeros for
reasonably large finite values $r$ in each of these figures. 

We find several general features of the boundary ${\cal B}$ for the $r \to
\infty$ limit of this family of graphs.  The first feature is that 
${\cal B}$ has support only for non-negative $Re(z)$ or equivalently,
non-negative $Re(q)$: 
\beq
z \in {\cal B} \Rightarrow Re(z) \ge 0 \quad {i.e.,} \quad Re(q) \ge 0
\label{bnonnegative}
\eeq
Indeed, the only place on the boundary ${\cal B}$ where $Re(z)$ vanishes is at
the origin $z=0$ itself: 
\beq
(z \in {\cal B} \quad {\rm and} \quad Re(z) = 0) \Rightarrow z = 0
\label{only0}
\eeq
Thus, in the vicinity of the point $z=0$, the curve comprising ${\cal B}$ bends
to the right as one increases $Im(z)$ above zero or decreases it below zero; we
have shown above that this curve crosses the point $z=0$ vertically. 
The second feature is that ${\cal B}$ has no multiple points \cite{mp}.  
The third feature concerns the
number of different connected components $N_{comp.}$ comprising ${\cal B}$. 
In the present case, this is simply related to the number of regions $N_{reg.}$
by the equation $N_{reg.}=N_{comp.}+1$. One might hope
that there would be a general mathematical theorem that would state the number
of different connected components $N_{comp.}$ 
of the solution set of a given polynomial equation
without requiring an explicit solution.  However this question, which is
related to the sixteenth Hilbert Problem, still remains unanswered 
\cite{hilbert16,amstran}.  One can make use of an upper bound on $N_{comp.}$
contained in the Harnack theorem, which is $N_{comp} \le g+1$, 
where $g$ denotes the genus of the algebraic curve \cite{alg}.  
For plane algebraic curves of the type relevant here, 
\beq
\sum_{m,n} c_{m,n} y_{_R}^m y_{_I}^n = 0
\label{yeqgeneral}
\eeq
with maximal degree $d$, where $y_{_R}=Re(y)$, $y_{_I} = Im(y)$, 
the genus is $g=(d-1)(d-2)/2$. Thus, for a given case, one
writes the degeneracy equation (\ref{yeqtkr}) or its equivalent in the $z$
plane, out into cartesian components.  
The case $k=2$ has already been discussed and leads to the equation 
$(y_{_R}- 1/3)^2 + y_{_I}^2 = (1/3)^2$, or equivalently, 
$(z_{_R}- 1/5)^2 + z_{_I}^2 = (1/5)^2$ where $z_{_R}=Re(z)$ and 
$z_{_I}=Im(z)$.  These equations have homogeneous degree \cite{alg} 
$d=2$, hence genus $g=0$, so Harnack's theorem yields the bound $N_{comp.}
\le 1$ which, together with the fact that $N_{comp.} \ge 1$ implies that
$N_{comp.}=1$, in agreement with the elementary solution of the explicit
equation above. However, for higher values of $k$, the Harnack upper bound is
not very restrictive.  For $k \ge 3$, eq. (\ref{yeqtkr}), when written out in 
cartesian components, is 
\beq
\Bigl [ (y_{_R}+1)^2 + y_{_I}^2 \Bigr ]F(y_{_R}, y_{_I})=0
\label{yeqk3} 
\eeq
where $F(y_{_R}, y_{_I})=3(y_{_R}^{2k-2} + y_{_I}^{2k-2})+$ lower order terms. 
Because the first factor is positive definite, eq. (\ref{yeqtkr}) thus reduces
to $F(y_{_R}, y_{_I})=0$, of homogeneous degree $d=2k-2$ and hence genus 
$g=(2k-3)(k-2)$.  The Harnack theorem then yields the upper bound 
$N_{comp.} \le 2k^2-7k+7$.  As an example, for $k=3,4,5,6$, this bound has
the respective values 4,11,22,37, while the actual values are 
$N_{comp.}=1,1,3,3$, and so forth for higher $k$. 

For both of the cases $k=3,4$, as was 
true for $k=2$, ${\cal B}$ divides the $z$ plane into two regions: (i) 
the region $R_1$ including the vicinity of the positive real axis contiguous
with the origin, $z=0$; and (ii) the region $R_2$ occupying the rest of the $z$
plane outside of $R_1$.  In the $z$ plane, the chromatic zeros tend to cluster
on the regions of the curve ${\cal B}$ in the ``northeast'' and ``southeast''
directions.  Indeed, as $k$ increases from 3 to 4, ${\cal B}(z)$ develops
protuberances in these ``northeast'' and ``southeast'' directions.  Since for a
given $r$ and $k$ the chromatic zeros are bounded, they avoid the origin of the
$z$ plane, as can be seen in the figures. Between
$k=4$ and $k=5$ there occurs a qualitative change in ${\cal B}$, namely, that,
whereas it consisted of a single component for $2 \le k \le 4$, it consists of 
three disconnected components for $5 \le k \le 7$, comprised of a
self-conjugate closed curve together with a complex-conjugate pair of closed
curves.  Hence, while ${\cal B}$ divides the $z$ plane into two regions for 
$2 \le k \le 4$, it divides this plane into four regions for $5 \le k \le 7$.
In a figurative language, one can think of the northeast and southeast 
bulges that are evident in the $k=4$ case as breaking off to form
the two separate closed curves that first appear for $k=5$.  We also observe
that the additional closed curves and associated disconnected regions appear 
to line up approximately along the vertical line with $Re(z)=1/2$.  It is
interesting to note that this line maps to unit circles in the $y$ and $q$
planes: 
\beq
z: \ Re(z)=\frac{1}{2}, \ Im(z) \ {\rm arbitrary} \Longleftrightarrow 
|y|=1 \quad \Longleftrightarrow \quad |q-1|=1
\label{lineimages}
\eeq
Thus, the disconnected phases that appear in the $z$ plane are clustered 
around the unit circle $|y|=1$ in the $y$ plane.  In the $q$ plane, portions of
the boundary ${\cal B}$ are also clustered around the circle $|q-1|=1$ while
one portion extends infinitely far from the origin.

\section{The Family $C_{\lowercase{k,r}} =
HEG_{\lowercase{k}-2}(\overline K_2 + C_{\lowercase{r}})$}

We obtain a further infinite family of graphs with noncompact ${\cal B}(q)$ 
by the same steps as in the previous section, but replacing the subgraph 
$T_r$ by the circuit subgraph $C_r$.  That is, we homeomorphically expand the
family $(K_p)_b + C_r$ for $r \ge 2$ by adding vertices on each of the 
bonds in the $C_r$ subgraph.  As before, we let $k$ be the number of vertices 
on $C_r$ between each pair of vertices that connect with the $K_p$ subgraph, 
including this pair.  We denote this family as 
\beq
C_{p,b,k,r} = HEG_{k-2}[(K_p)_b + C_r]
\label{cpbkr}
\eeq
where as above, $r \ge 2$ and $1 \le b \le p-1$.
For a given set of parameters $p$, $k$, and $r$, the number of vertices in this
family is
\beq
v(C_{p,b,k,r}) = r(k-1)+p
\label{vcpbkr}
\eeq
For the lowest value, $k=2$,
\beq
C_{p,b,k=2,r} = (K_p)_b + C_r
\label{cpbkeq2r}
\eeq
which we studied previously \cite{wa}; accordingly, we concentrate here
on the cases $k \ge 3$.  It suffices for our present purposes to consider the
simplest nontrivial case $p=2$ and hence $b=1$, for which
\beq
(K_2)_{b=1} + C_r = \overline K_2 + C_r
\label{c2bartr}
\eeq
As before, a short notation is useful: 
\beq
C_{k,r} = C_{p=2,b=1,k,r} = HEG_{k-2}(\overline K_2 + C_r)
\label{ckrdef}
\eeq
An illustration of a graph of this type is given in Fig. \ref{vhegraph1}(b).
The chromatic number and girth are the same as for $T_{k,r}$, given by the
right-hand sides of eqs. (\ref{chitkr}) and (\ref{girthtkr}). 
By the deletion-contraction theorem, we find the recursion relation
\beq
P(C_{k,r},q) = D_k P(T_{k,r},q) + (-1)^{k-1}P(C_{k,r-1},q)
\label{ckrrecursion}
\eeq
For the lowest value of $r$, namely, $r=2$, we have 
\beq
P(C_{k,2},q) = \Bigl [ D_k + (-1)^{k-1} \Bigr ] P(T_{k,2},q) + 
(-1)^k (D_3)^2 P(C_k,q) 
\label{ckr2}
\eeq
Solving the recursion relation (\ref{ckrrecursion}) with (\ref{ckr2}), we 
calculate the chromatic polynomial.  As before, we express this in the form of 
eq. (\ref{pgsum}).  (In order to render the polynomial property manifest, one
divides through by the factors in the denominator, thereby generating a series,
using the identity $(x^m-1)/(x-1)=\sum_{j=0}^{m-1}x^j$.)  We get
\beqs
P(C_{k,r},q) & = &
D_k \Biggl [ A_k a_2 \Bigl ( a_2 + (-1)^{k-1}\Bigr )
\biggl [ \frac{a_2^{r-2}-(-1)^{(k-1)r}}{a_2^2-1} \biggr ] + \cr\cr\cr
& & q(q-1)a_1^2\Bigl (a_1 + (-1)^{k-1} \Bigr )
\biggl [ \frac{a_1^{r-2}-(-1)^{(k-1)r}}{a_1^2-1} \biggr ] \Biggr ] +
(-1)^{(k-1)r}P(C_{k,2},q)
\label{pckr}
\eeqs
where $A_k$ was defined in eq. (\ref{ak}). 
$P(C_{k,r},q)$ has the general factors $q(q-1)$ for $k$ odd and
$q(q-1)(q-2)$ for $k$ even.  $P(C_{k,r},q)$ has the form of 
eq. (\ref{pgsum}) with $a_1$ and $a_2$ as given in eqs. (\ref{a1tkr}) and 
(\ref{a2tkr}), together with 
\beq
a_3 = 1
\label{a3ckr}
\eeq
For large $|q|$ (small $|z|$), 
the terms $a_1$ and $a_2$ have larger magnitudes than $a_3$, so that in the
limit $r \to \infty$ (with $k$ fixed \cite{lknote}) 
the boundary ${\cal B}$ is determined by the degeneracy equation 
$|a_1|=|a_2|$, and the same analysis goes through as before for
the family $HEG_{k-2}(\overline K_2 + T_r)$, with the same conclusions that
${\cal B}$ is noncompact in the $q$ plane, passing through the point $z=y=0$
vertically.  For the lowest case $k=2$, i.e., $\overline K_2 + C_r$, 
${\cal B}$ consists of three regions, as shown in Fig. \ref{bipzplane}): 
$R_1$ including the real axis in the interval $0 \le z \le z_c$ with 
$z_c=1/3$ as given by eq. (\ref{qckpcr}); $R_3$ centered around $z=2/5$ and
occupying the interval $z_c \le z \le 1/2$ on the real axis; and $R_2$
occupying the complement of the $z$ plane extending outward to $|z| \to
\infty$.  This region diagram may be contrasted with
the simpler one for $\overline K_2 + T_r$, which was just a circle, separating
the $z$ plane into two regions. 

\begin{figure}
\vspace{-4cm}
\centering
\leavevmode
\epsfxsize=3.0in
\begin{center}
\leavevmode
\epsffile{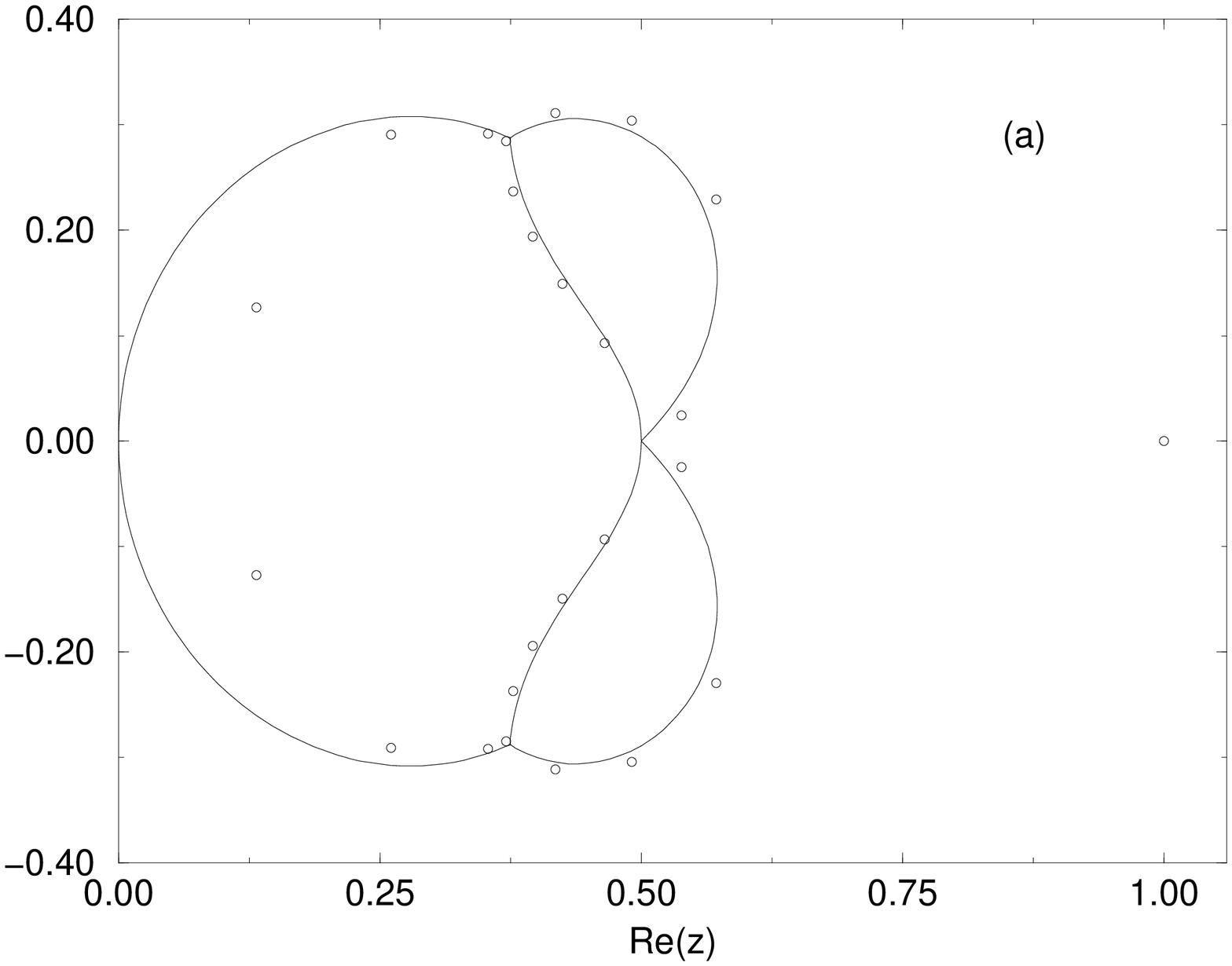}
\end{center}
\vspace{-2cm}
\begin{center}
\leavevmode
\epsfxsize=3.0in
\epsffile{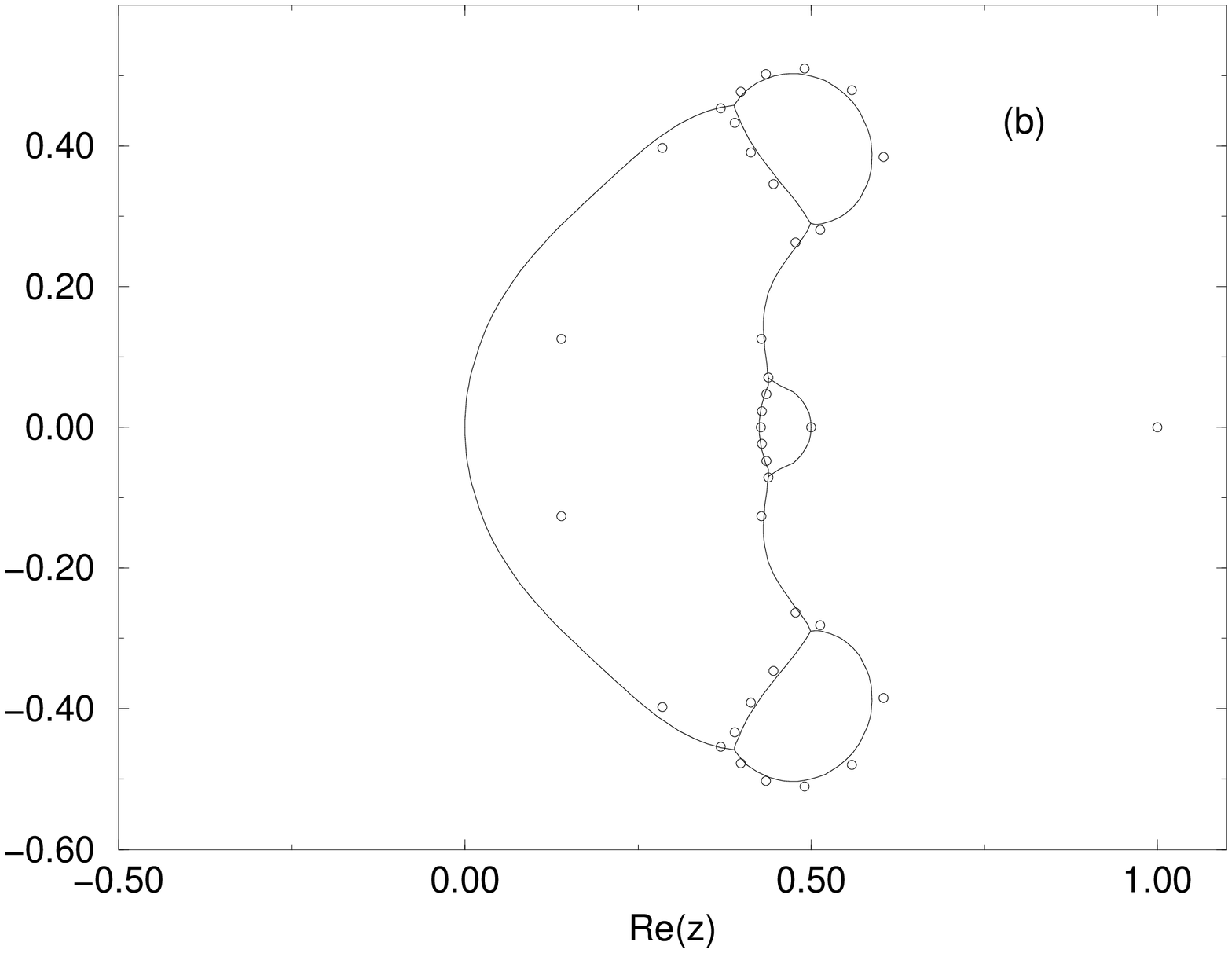}
\end{center}
\vspace{-2cm}
\caption{\footnotesize{Boundary ${\cal B}$ in the $z=1/q$ plane for the $r \to
\infty$ limit of the family of graphs $HEG_{k-2}(\overline K_2 + C_r)$
with $k=$ (a) 3 (b) 4. Chromatic zeros for $r=12$ are shown for 
comparison.}}
\label{boundarye2hvkcr1}
\end{figure}

\begin{figure}
\vspace{-4cm}
\centering
\leavevmode
\epsfxsize=3.0in
\begin{center}
\leavevmode
\epsffile{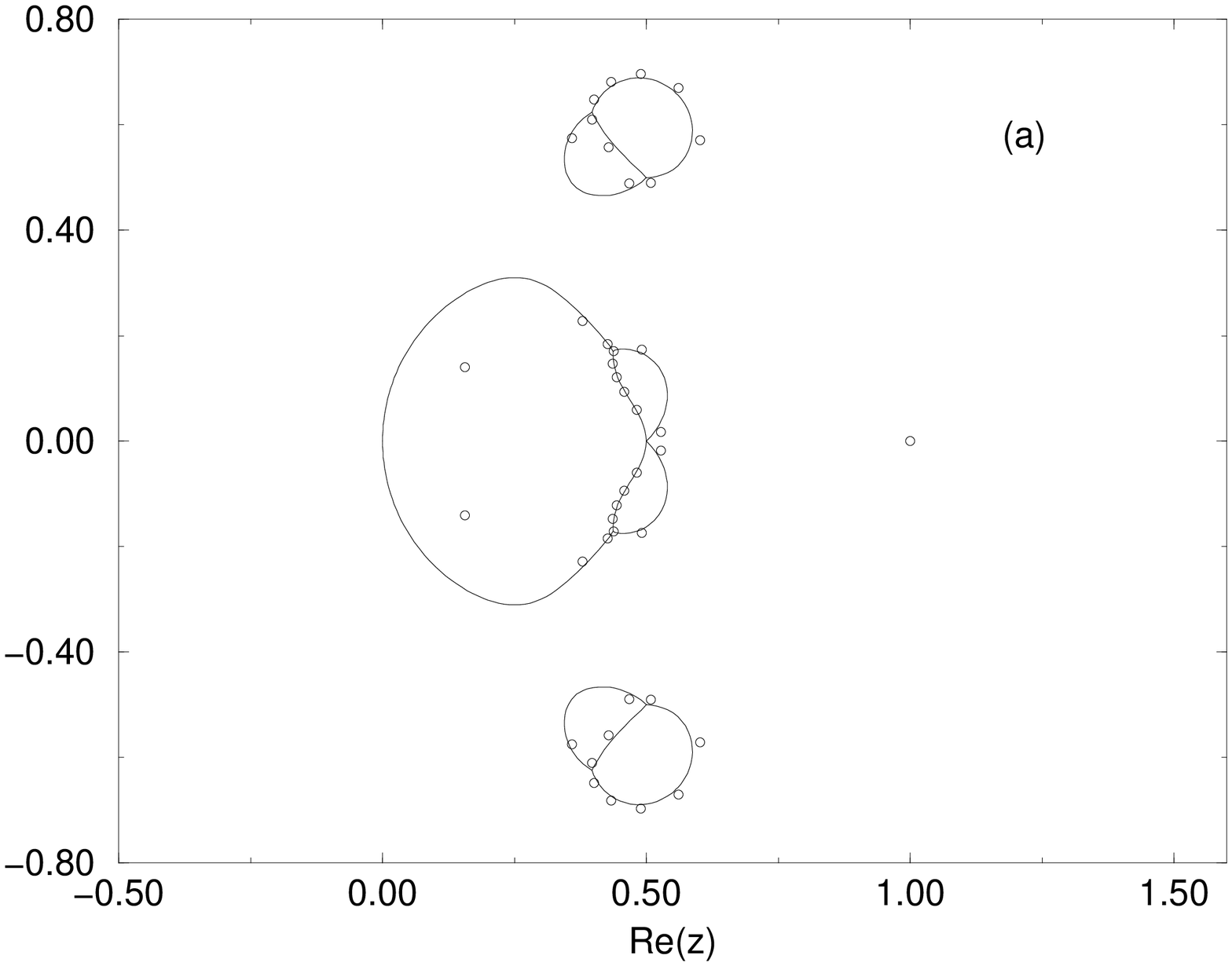}
\end{center}
\vspace{-2cm}
\begin{center}
\leavevmode
\epsfxsize=3.0in
\epsffile{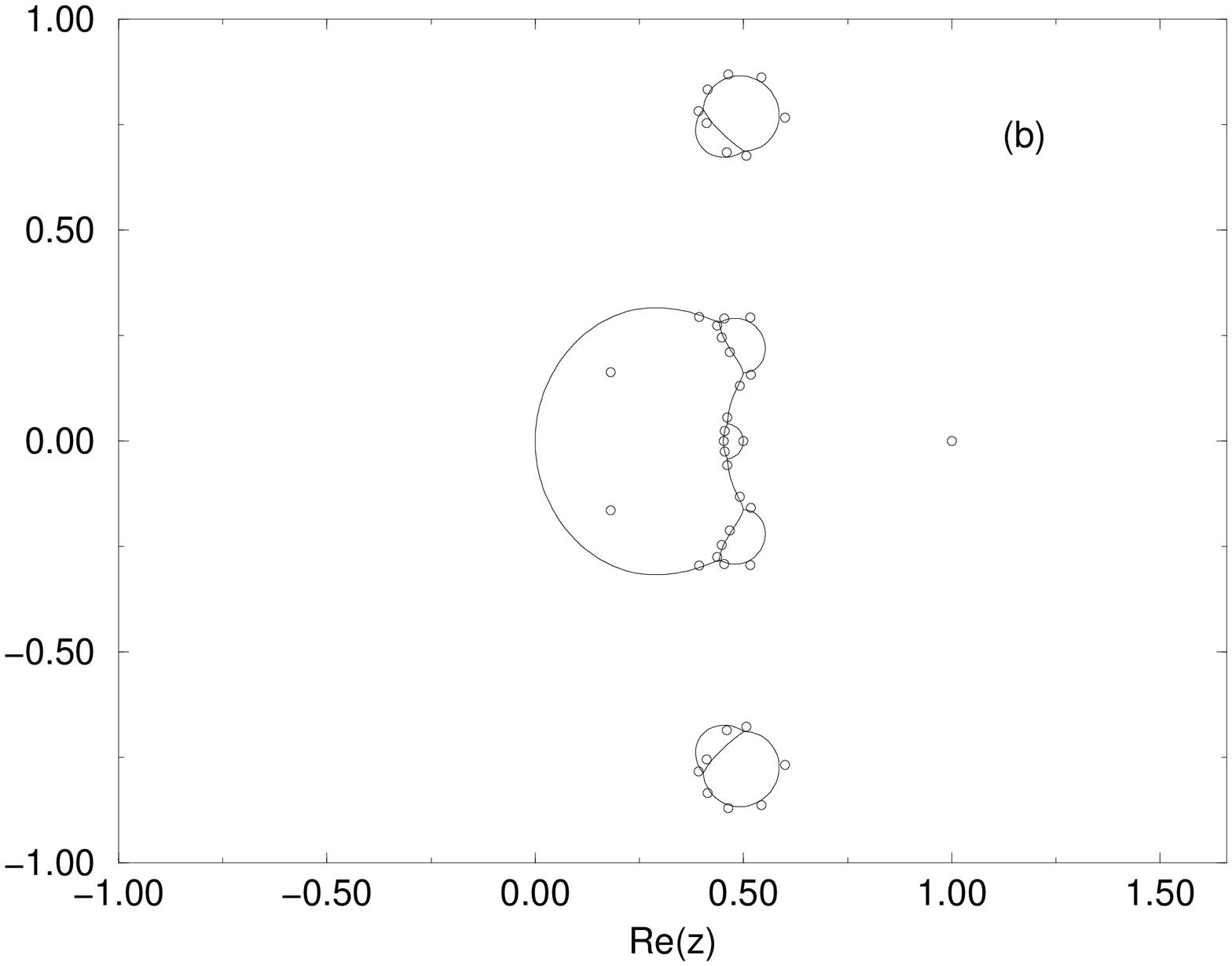}
\end{center}
\vspace{-2cm}
\caption{\footnotesize{As in Fig. \ref{boundarye2hvkcr1} for $k=$ (a) 5 (b) 6.
Chromatic zeros for $r=$ (a) 10 (b) 8 are shown for comparison.}}
\label{boundarye2hvkcr2}
\end{figure}

In Figs. \ref{boundarye2hvkcr1} and \ref{boundarye2hvkcr2} we 
show the boundary ${\cal B}$ in the $z$ plane for $3 \le k \le 6$ 
\cite{thesis}. 
Again, for comparison, we show chromatic zeros for reasonably large finite 
values $r$ in each of these figures.  The property that ${\cal B}$ passes
through $z=0$ vertically is evident in these figures, as are the properties
(\ref{bnonnegative}) and (\ref{only0}).  For each value of $k$,
the region $R_1$ is the one occupying the non-negative interval of the $z$ axis
adjacent to the origin.  The region $R_2$ occupies 
the rest of the $z$ plane for $2 \le k \le 4$, and
the rest of the $z$ plane except for the interiors of the other closed loops
for $k \ge 5$.  For odd $k$, there are two regions on the real $z$ axis: 
$R_1$ for $0 \le z \le 1/2$ and $R_2$ for $z < 0$ and $z > 1/2$, while for even
$k \ge 4$ there are these phases and, in addition, a third ``pocket'' phase
$R_3$ whose right-hand boundary with $R_2$ occurs at $z=1/2$ and whose 
left-hand boundary with $R_1$ occurs at the point $z=z_c$ slightly less than 
1/2 (see further below). 

As was the case for 
$\lim_{r \to \infty}[HEG_{k-2}(\overline K_2 + T_r)]$, we find 
\beq
W([\lim_{r \to \infty} C_{k,r}],q) = (a_1)^{1/(k-1)} = (D_{k+1})^{1/(k-1)}
 \quad {\rm for} \quad q \in R_1
\label{wckr_r1}
\eeq
and 
\beq
|W([\lim_{r \to \infty} C_{k,r}],q)| = |a_2|^{1/(k-1)} = |D_3D_k +
(-1)^{k-1}|^{1/(k-1)} \quad {\rm for} \quad q \in R_2
\label{wckr_r2}
\eeq
Hence for $k=3$ the equalities (\ref{tkrwq2region1}) and 
(\ref{tkrwq2region2}) hold, with $T_{k,r}$ replaced by $C_{k,r}$. 
For $k \ge 3$, additional regions appear as ``outgrowths'' on the right-hand
boundary of $R_1$ in the $z$ plane, bounded on the left by $R_1$ and on the
right by $R_2$.  We label these regions generically as $R_{1-2;j}$, where $j
\ge 3$ indexes the region.  For example, there are two such regions for
$k=3$, which are complex-conjugates of each other, and there are three such
regions for $k=4$, consisting of a complex-conjugate pair and the region $R_3$
that includes an interval of the real $z$ axis between $z=0.42495$ and 
$z=1/2$.  For
arbitrary $k \ge 3$, in these outgrowth regions lying between $R_1$ and $R_2$, 
the leading term 
in the limit $r \to \infty$ is the last one in eq. (\ref{pckr}), 
viz., $(-1)^{(k-1)r}P(C_{k,2},q)$, so that 
\beq
|W([\lim_{r \to \infty} C_{k,r}],q)| = 1 
\quad {\rm for} \quad q \in R_{1-2,j}
\label{wckr_rj}
\eeq
For $k \ge 5$, additional sets of regions occur that are not connected with 
$R_1$; these consist of complex-conjugate pairs, lying near to the vertical 
line in the $z$ plane with $Re(z) = 1/2$.  
For the cases that we have studied, each of these disconnected sets of regions
actually consists of two: for the ones with $Im(z) > 0$, one part occupying 
a ``northeast'' position and the other a ``southwest'' position, as is evident
in the figures.  We label these regions disconnected from $R_1$ as 
$R_{disc,NE;j}$, $R_{disc,SW;j}$, and their complex-conjugates as 
$R_{disc,NE;j}^*$, $R_{disc,SW;j}^*$
. We find that
\beq
|W([\lim_{r \to \infty} C_{k,r}],q)| = |a_1|^{1/(k-1)} = |D_{k+1}|^{1/(k-1)} 
\quad {\rm for} \quad q \in R_{disc,SW;j}, R_{disc,SW;j}^*
\label{wckr_rdisc_sw}
\eeq
\beq
|W([\lim_{r \to \infty} C_{k,r}],q)| = 1 
\quad {\rm for} \quad q \in R_{disc,NE;j}, R_{disc,NE;j}^*
\label{wckr_rdisc_ne}
\eeq
As was noted before in the case of the $r \to \infty$ limit of the family
$T_{k,r}$, this clustering of the disconnected phases along the 
vertical line $Re(z)=1/2$ is equivalent to their clustering around the unit
circle in the $y$ plane.

For odd $k$, ${\cal B}$ crosses the real $z$ axis away from $z=0$ 
(equivalently the real $q$ axis) once, at $z=1/2$ (i.e., $q=2$).  
For even $k$, ${\cal B}$ crosses the real $z$ axis at two points away from 
the origin $z=0$: (i) at $z=1/2$, and (ii) at a value of $z$ that increases 
monotonically from $z=1/3$ for $k=2$, approaching $z=1/2$ as 
$k \to \infty$ through even integers.  Hence, taking into account that 
\beqs
\chi(C_{k,r}) & = & 2 \quad {\rm for} \quad k \quad {\rm odd} \cr
              & = & 3 \quad {\rm for} \quad k \quad {\rm even}
\label{chickr}
\eeqs
we have 
\beq
q_c = 2 = \chi \quad {\rm for \ odd} \quad k
\label{qce2ckr_kodd}
\eeq
while 
\beq 
2 < q_c \le 3 = \chi \quad {\rm for \ even} \quad k
\label{qce2ckr_keven}
\eeq
Equations (\ref{qce2ckr_kodd}) and (\ref{qce2ckr_keven}) may be compared with
the analogous eqs. (\ref{qce2tkr_kodd}) and (\ref{qce2tkr_keven}) for 
the family $T_{k,r}= HEG_{k-2}(\overline K_2 + T_r)$. 
To derive the above results, we observe first that for 
the case of odd $k$, this follows from the same analysis as applied to the
$r \to \infty$ limit of the family $ HEG_{k-2}(\overline K_2 + T_r)$ in the
previous section, since the crossing is determined by the degeneracy condition
$|a_1|=|a_2|$. For even $k$, the crossing point at the smaller value of $z$ is
determined by the degeneracy condition of leading terms
\beq
{\cal B}(R_1,R_3) \ : \quad |a_1|=|a_3|=1
\label{br1r3}
\eeq
To put this into a form that can be used for the analysis of our region 
diagrams in the $z$ plane, Figs. 
\ref{boundarye2hvkcr1} and \ref{boundarye2hvkcr2}, we 
multiply both sides of eq. (\ref{br1r3}) by $|q(q-1)|=|(a+1)a|$ 
(the spurious solutions at $q=0$ and $q=1$ thereby introduced are understood 
to be discarded).  The resulting equation is most
simply expressed in the $y$ plane, as
$|1-y^k|=|y^{k-1}(y+1)|$, i.e., 
\beq
{\cal B}(R_1,R_3): \quad 
\rho^{2k-2} + 2\rho^k \Bigl (\rho^{k-1}\cos\beta + \cos(k\beta)\Bigr ) -1 = 0
\label{br1r3polar}
\eeq
For $\beta=0$, this equation has a single acceptable (real positive) root for 
$\rho=y_c$ that increases monotonically from 1/2 (i.e., $z_c=1/3$) for $k=2$ 
through 0.738984 ($z=0.424951$) for $k=4$, toward $\rho=y_c=1$ ($z_c=1/2$) 
as $k$ goes to infinity through even values; 
equivalently, $q_c$ decreases from 
3 at $k=2$ toward 2 as $k \to \infty$ through even values.  Some explicit 
values are $q_c=2.35321$, 2.21486, 2.15442 for $k=4,6,8$. 
The crossing point at the larger value of $z$ (for even $k$), 
viz., $z=1/2$, is determined by the degeneracy condition of leading terms 
$|a_2|=|a_3|$.  Multiplying by $|q(q-1)|$ and reexpressing this in terms of $y$
as before, we obtain $|1-y+2(-1)^{k-1}y^k|=|(1+y)y^{k-1}|$, i.e., 
\beqs
{\cal B}(R_2,R_3): \quad & & 
3\rho^{2k}-\rho^{2k-2}+\rho^2+1 - 2\rho(\rho^{2k-2}+1)\cos\beta \cr\cr
& & - 4\rho^k\Bigl [ \cos(k\beta) - \rho\cos((k-1)\beta) \Bigr ] = 0
\label{br2r3polar}
\eeqs
For $\beta=0$, this equation always has a root at $\rho=y=1$, (i.e., $z=1/2$,
$q=2$).  (It also has a real positive root at a value of $\rho < 1$ 
(i.e., $z < 1/2$), but $a_2$ is not a leading term in this region, so that
this root is irrelevant.) 

The number of multiple points on ${\cal B}$ is 
\beq
N_{m.p.}=2(k-1) \ , \quad {\rm for \ even} \quad k
\label{nmpe2hvcr_evenk}
\eeq
comprised of $k-1$ complex-conjugate pairs, and 
\beq
N_{m.p.}=2k-3
\label{nmpe2hvcr_oddk}  \ , \quad {\rm for \ odd} \quad k
\eeq
consisting of a real one at $z=1/2$, i.e., 
$q=2$, together with $k-2$ complex-conjugate pairs. 

It is interesting to compare the boundaries ${\cal B}$ and associated region
diagrams for the $r \to \infty$ limits of the two families of graphs 
$HEG_{k-2}(\overline K_2 + T_r)$ and $HEG_{k-2}(\overline K_2 + C_r)$ for a 
given value of $k$.  We have done this above for $k=2$.  For $k \ge 3$, one
sees that the diagrams for $HEG_{k-2}(\overline K_2 + C_r)$ look somewhat
similar to those for $HEG_{k-2}(\overline K_2 + T_r)$ with the differences that
(i) there are additional ``outgrowth'' regions contiguous to the right-hand
side of $R_1$; (ii) the additional disconnected regions that appear for $k
\ge 5$ are themselves composed of additional regions; (iii) ${\cal B}$
contains multiple points; and (iv) for even $k$ there are three, rather than
just two regions along the real $z$ axis.  Owing to the presence of the
multiple points on ${\cal B}$ (i.e. singular points, in the sense of algebraic
geometry), the analysis of bounds on the number of disconnected components,
$N_{comp.}$ is more difficult than for algebraic curves without singular
points, such as the curves ${\cal B}$ for $\lim_{r \to \infty}T_{k,r}$. 

\pagebreak

\pagebreak

\section{The Family $S_{\lowercase{k,r}} =
HEG_{\lowercase{k}-2}(\overline K_3 + T_{\lowercase{r}})$}

To illustrate one infinite family of graphs with noncompact ${\cal B}$ of the
type $HEG_{k-2}(\overline K_p + G_r)$ with $p>2$, we discuss an infinite
family constructed by homeomorphic expansion of the bonds of the subgraph
$G_r=T_r$ in the graph $(\overline K_3 + T_r)$.  The beginning family, 
$\overline K_3 + G_r$, is of type (\ref{epg}) (and can be obtained by 
complete removal of all three of the bonds in the $K_3$ subgraph in 
$K_3 + G_r$). We denote this family as 
\beq
S_{\lowercase{k,r}} = HEG_{k-2}(\overline K_3 + T_r)
\label{skr}
\eeq
The number of vertices is given by
\beq
v(S_{k,r})=(r-1)(k-1)+4
\label{vskr}
\eeq
An example of a graph of this type is shown in Fig. \ref{vhegraph1}(c).
The chromatic number and girth are the same as those of the graph $T_{k,r}$,
given in eqs. (\ref{chitkr}) and (\ref{girthtkr}). 
By the deletion-contraction theorem, we find the recursion relation
\beq
P(S_{k,r},q) = \Bigl [ (q-3)D_k + (-1)^{k-1} \Bigr ] P(S_{k,r-1},q) +
3D_k P(T_{k,r-1},q) - q(q-1)D_k (D_{k+1})^{r-2}
\label{skrrecursion}
\eeq
Solving this, we obtain the chromatic polynomial
\beqs
P(S_{k,r},q) & = & 3 \biggl [ q(q-1)D_{k+1}-P(T_{k,2},q) \biggr ] 
\biggl [ a_3^{r-2}-a_2^{r-2} \biggr ]
+q(q-1)a_1 \biggl [a_1^{r-2}-a_3^{r-2} \biggr ] \cr\cr
 & & + a_3^{r-2}P(S_{k,2},q)
\label{pskr}
\eeqs
where $a_1$ and $a_2$ were given in eqs. (\ref{a1tkr}) and (\ref{a2tkr}),
\beq
a_3=(q-3) D_k + (-1)^{k-1} \ , 
\label{a3skr}
\eeq
and
\beq
P(S_{k,2},q)=(q-2)P(T_{k,2},q)+q(q-1)^3D_{k-1}
\label{pskr2}
\eeq
In the limit $r \to \infty$ with $k$ fixed \cite{lknote}, the locus 
${\cal B}$ is determined by the degeneracy of magnitudes of leading
terms 
\beq
|a_1|=|a_3|
\label{degeneqskr}
\eeq
($a_2$ is never a leading term in this case). The degeneracy equation in the 
$y$ variable takes the form
\beq
|1-2y+3(-1)^{k+1}y^k| = |1+(-1)^{k+1}y^k|
\label{yeqskr}
\eeq
Since $y=0$ is a solution of this equation, ${\cal B}$ is noncompact in the
$q$ plane, passing through $y=z=0$. In polar coordinates, with
$y=\rho e^{i\beta}$, eq. (\ref{yeqskr}) yields
\beq
\rho\Bigl [2\rho^{2k-1} + \rho - \cos \beta + (-1)^{k-1}\rho^{k-1}
\biggl \{ \cos(k\beta) - 3\rho \cos((k-1)\beta) \biggr \} \Bigr ] = 0
\label{yeqskrpolar}
\eeq
As $\rho \to 0$, it follows that $\cos\beta=0$, i.e. $\beta = \pm \pi/2$, so
that ${\cal B}$ approaches $y=z=0$ vertically.

To calculate the point at which the boundary ${\cal B}$ crosses the real $q$,
or equivalently, $y$ or $z$ axes, we set $\beta=0$ in
eq. (\ref{yeqskrpolar}); for $\rho \ne 0$ this gives
\beq
2\rho^{2k-1} + \rho - 1 + (-1)^{k-1}\rho^{k-1}(1-3\rho)=0
\label{yeqskrpolarpositivey}
\eeq
For both even and odd $k$, eq. (\ref{yeqskrpolarpositivey}) has only one
acceptable (real positive) root for $\rho$, which is thus $y_c$. 
For odd $k$ this root is $y_c=1$
(equivalently, $z_c=1/2$, $q_c=2$). For even $k$, the real positive crossing 
point increases monotonically from $y_c=0.647799$ ($z_c=0.39313$) for $k=4$ 
to $y_c=1$ ($z_c=1/2$) as $k$ goes to infinity through even values; i.e., 
$q_c$ decreases monotonically from 2.54369 to 2 over the same range. 

\begin{figure}
\vspace{-4cm}
\centering
\leavevmode
\epsfxsize=3.0in
\begin{center}
\leavevmode
\epsffile{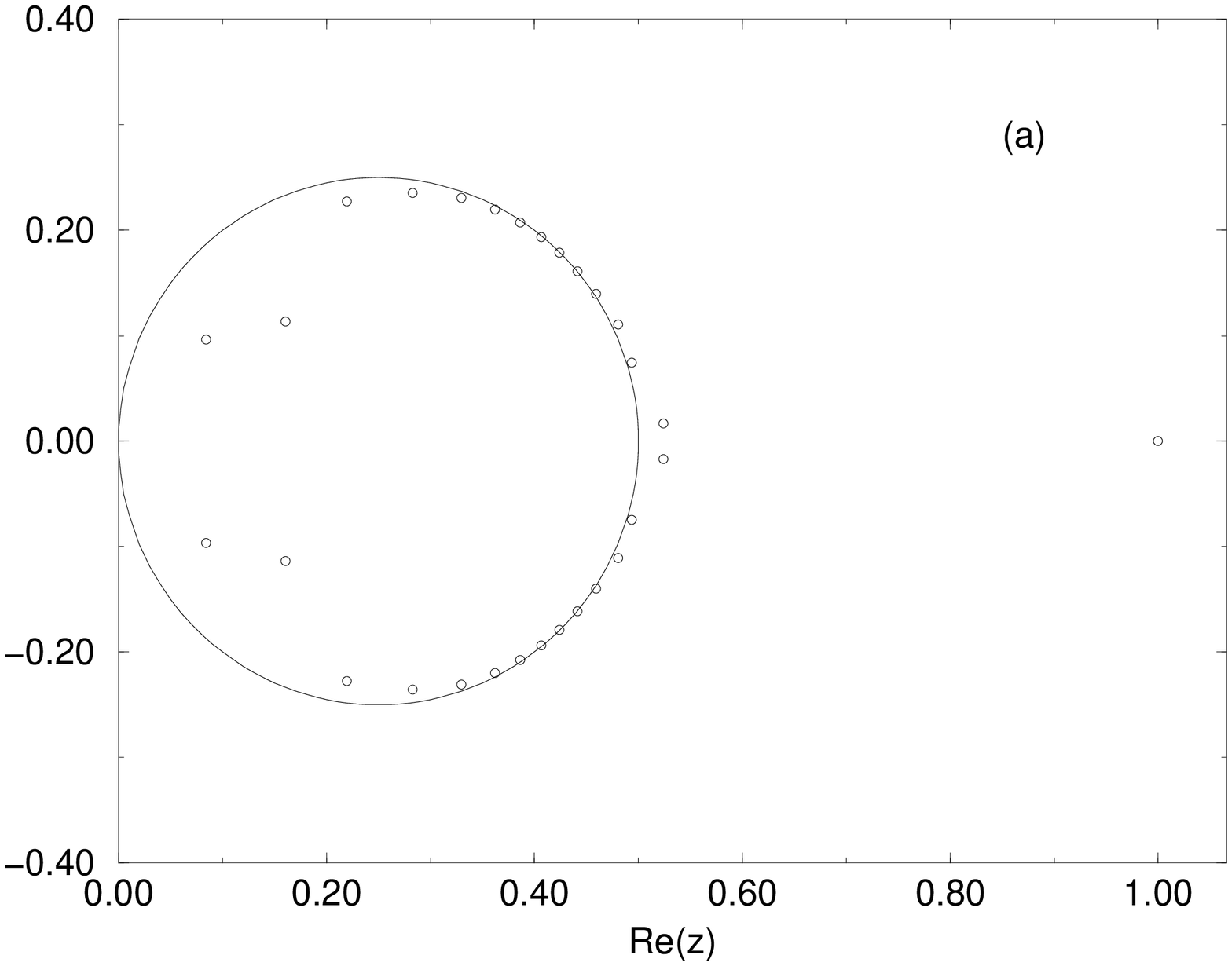}
\end{center}
\vspace{-2cm}
\begin{center}
\leavevmode
\epsfxsize=3.0in
\epsffile{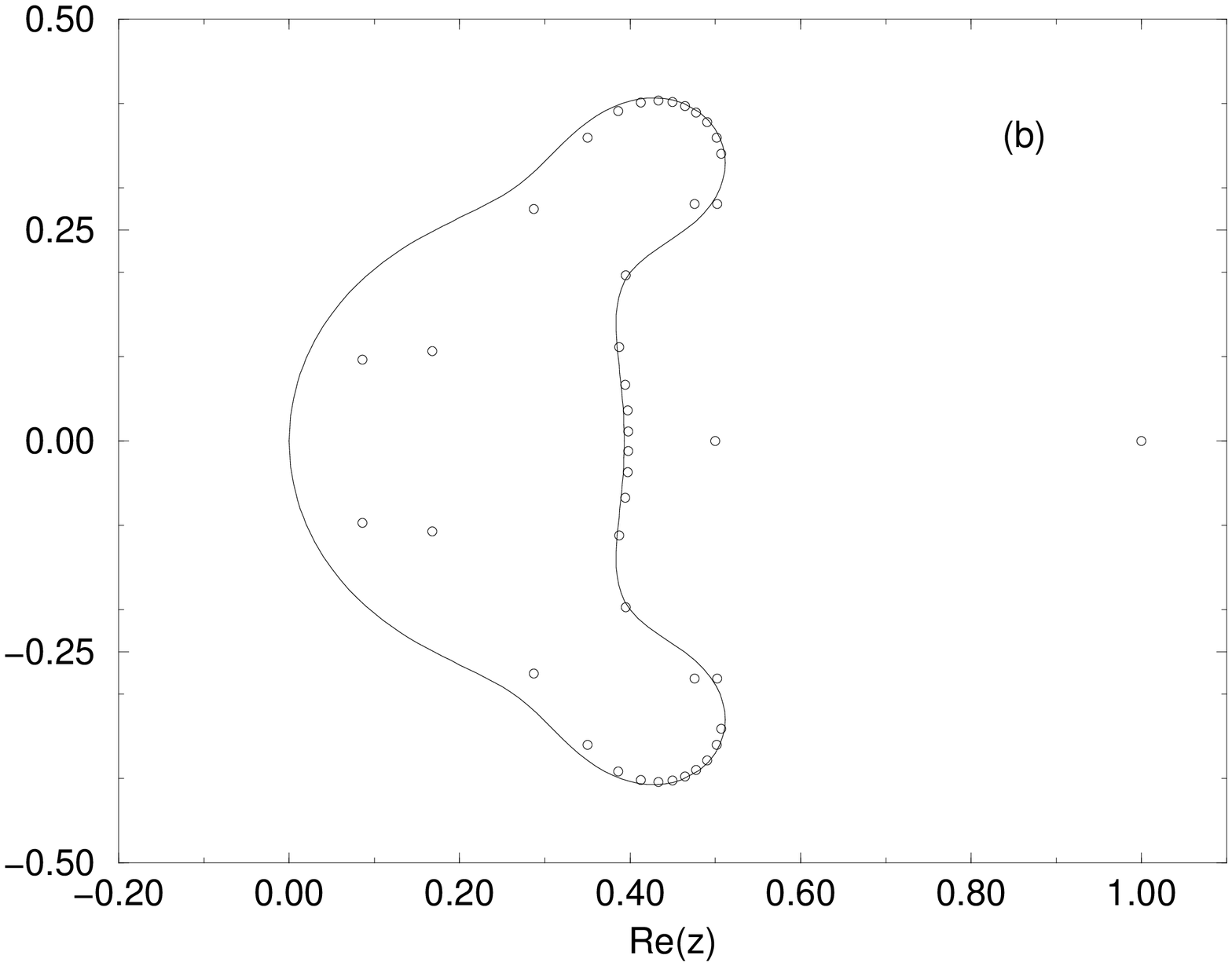}
\end{center}
\vspace{-2cm}
\caption{\footnotesize{Boundary ${\cal B}$ in the $z=1/q$ plane for the $r \to
\infty$ limit of the family of graphs $HEG_{k-2}(\overline K_3 + T_r)$
with $k=$ (a) 3 (b) 4. Chromatic zeros for $r=14$ are shown for
comparison.}}
\label{boundarye3hvktr1}
\end{figure}

In Fig. \ref{boundarye3hvktr1} we show the 
respective boundaries ${\cal B}$ in the $z$ plane for $k=3,4$ \cite{thesis}.  
Note that for $k=3$, ${\cal B}$ is close to being, but is not, a circle.   
Again, for comparison, we
show chromatic zeros for reasonably large finite values of $r$ in each of these
figures.  As is evident from Fig. \ref{boundarye3hvktr1} and from the figures
for higher values of $k$ \cite{thesis}, 
${\cal B}$ for $\lim_{r \to \infty} S_{k,r}$ displays some similarities with 
that for $\lim_{r \to \infty} T_{k,r}$.  Hence we do not show these figures for
higher values of $k$ here.  We remark, however, that there are some 
differences; for example, the number of connected components, $N_{comp.}$ is
equal to the respective values 1,1,3,3,3,5 for $3 \le k \le 8$ for 
$\lim_{r \to \infty} T_{k,r}$ but is equal to 1,1,3,3,5,5 for the present case,
$\lim_{r \to \infty} S_{k,r}$ with $3 \le k \le 8$.  As was true with 
$N_{comp.}$ for $\lim_{r \to \infty}T_{k,r}$, for $k \ge 4$, 
the values of $N_{comp.}$ are
only weakly bounded by the Harnack theorem. 

For each value of $k$,
the region $R_1$ is the one occupying the non-negative interval of the $z$ axis
adjacent to the origin.  The region $R_2$ occupies
the rest of the $z$ plane for $2 \le k \le 4$, and
the rest of the $z$ plane except for the interiors of the other closed loops
for $k \ge 5$. We find
\beq
W(\lim_{r \to \infty} S_{k,r},q) = (a_1)^{1/(k-1)} = (D_{k+1})^{1/(k-1)}
 \quad {\rm for} \quad q \in R_1
\label{wskr_r1}
\eeq
and
\beq
|W([\lim_{r \to \infty} S_{k,r}],q)| = |a_3|^{1/(k-1)} = |(q-3)D_k +
(-1)^{k-1}|^{1/(k-1)} \quad {\rm for} \quad q \in R_2
\label{wskr_r2}
\eeq
For $k=5,6$ ($k=7,8$) there is one pair (are two pairs) of complex-conjugate
closed loops disconnected from $R_1$, which we denote $R_{disc,j}$ and
$R_{disc,j}^*$. In these regions we find
\beq
|W([\lim_{r \to \infty} S_{k,r}],q)| = (a_1)^{1/(k-1)} = (D_{k+1})^{1/(k-1)}
 \quad {\rm for} \quad q \in R_{disc,j}, R_{disc,j}^*
\eeq
The feature observed in the $r \to \infty$ limits of the families 
$T_{k,r}$ and $C_{k,r}$ that the
disconnected regions $R_{disc,j}$ and $R_{disc,j}^*$ tend to lie along the
line $z=1/2$ is also present in this case.

\pagebreak

\section{General Condition for Noncompact ${\cal B}$} 

   The key ingredient 
to construct families of graphs with noncompact $W$ boundaries ${\cal B}$ in
the $q$ plane and resultant reduced functions $W_{red.}(\{G \},q)$ that are
nonanalytic at $z=1/q=0$ is to produce a chromatic polynomial with
the feature that there are two leading terms with a degeneracy condition
(\ref{mageq}) that has a solution at $z=0$ (cf. eq. (\ref{mageqz})). 
A general statement of this condition was given as the theorem of Section IV of
Ref. \cite{wa}.  From our studies in the present paper, we can add some
geometrical comments to this algebraic condition.

An important property of our families of graphs with noncompact $W$ boundaries 
is that none of these families is a regular lattice graph.  This is in accord
with the derivation of the large--$q$ expansion for regular lattices
\cite{nagle,baker}.  (Of course, the property that $\{G\}$ is a regular
lattice is not a necessary condition for the associated boundary ${\cal B}$ to
be compact in the $q$ plane and hence $W_{red.}(\{G\},q)$ to be
analytic at $z=0$.)  A basic feature of a regular lattice graph is that (except
for boundary vertices, which yield a negligible effect in the thermodynamic
limit and are absent if one uses periodic boundary conditions) all vertices
have the same coordination number (= degree $\Delta$ in usual graph theory 
terminology).  A generalization of this is encountered in duals of 
Archimedean lattices, such as the $[4 \cdot 8^2]$ (union-jack) lattice, where 
the vertices fall into a finite number of sets with different coordination 
numbers (e.g. $\Delta=4$ and $\Delta=8$ for the $[4 \cdot 8^2]$ lattice).  In
both of these classes of lattices, there are no vertices with infinite degree.

In contrast, a property of the families of graphs that we have studied with 
noncompact loci ${\cal B}(q)$ in the limit $r \to \infty$ is that in this limit
they all contain an infinite number of different, non-overlapping 
(and non-self-intersecting) circuits,
each of which passes through at least two fixed, nonadjacent vertices.  This
immediately implies that these aforementioned nonadjacent vertices have 
degrees $\Delta$ that go to infinity in this limit.  
In the families that we constructed earlier \cite{wa} 
and some of the additional families discussed in sections II and
III of the present work, this property was produced by starting with a 
family of the form $K_p + G_r$ and removing one or more bonds from the $K_p$
subgraph, thereby rendering two or more vertices that were originally adjacent
no longer adjacent.  As we showed in Section IV, it can also be produced by
homeomorphically adding one or more vertices to a bond in the $K_p$ subgraph,
thereby again rendering two vertices that were originally adjacent no longer
so.  It should be mentioned that the condition of having two or more vertices 
with $\lim_{r \to \infty} \Delta(r) = \infty$ is not, by itself, sufficient to
produce a noncompact ${\cal B}(q)$.  For example, 
consider the family of ``$p$-wheels'', 
\beq
(Wh)_n^{(p)} = K_p + C_r
\label{wheel}
\eeq
(where $n=p+r$ is the number of vertices) that we constructed in Ref. 
\cite{wc}.  The degree of each of the $p$ vertices in the 
$K_p$ subgraph of the graph $K_p + C_r$ is $\Delta = p-1+r$, so that in the
limit $r \to \infty$ (with $p$ fixed), this degree $\Delta \to \infty$.
However, the corresponding boundary ${\cal B}$ is compact; specifically, we
showed that it is the unit circle $|q-(p+1)|=1$ \cite{wc}.  
Furthermore, we observe that the presence of non-adjacent vertices with 
degrees $\Delta$ such that $\lim_{r \to \infty}\Delta = \infty$ is also 
not, by itself sufficient to guarantee that ${\cal B}(q)$ is noncompact. 
This is easily seen by considering, for example, two $p=1$ $p$-wheel graphs,
$(Wh)_n^{(1)}$ whose central vertices are connected by a tree graph 
containing at least two bonds.  As $n \to \infty$, the degrees $\Delta$ of the
central vertices go to infinity, but again the resultant $W$ boundary 
${\cal B}$ is compact; indeed, it is the same as for a single $p=1$ $p$-wheel,
viz., $|q-2|=1$.  Thus, a necessary feature is observed to be the presence, in
the limit as the number of vertices of the graph goes to infinity, of an 
infinite number of different, non-overlapping (non-self-intersecting) 
circuits, each of which pass
through the two or more nonadjacent vertices.  

\section{Conclusions}

   In this paper we have further explored a fundamental problem in statistical
mechanics -- nonzero ground state entropy -- using as a theoretical laboratory
the $q$-state Potts antiferromagnet. We have presented a number of exact 
calculations of the zero-temperature partition function $Z(G,q,T=0)$ or
equivalently the chromatic polynomial $P(G,q)$, and the corresponding limiting
function representing the ground-state degeneracy, $W(\{G\},q)$, for this model
on various families of graphs $G$ for which the boundary ${\cal B}$ of 
regions of analyticity of $W$ in the complex $q$ plane is noncompact, passing 
through $z=1/q=0$.  The study of these graphs thus gives insight into the 
conditions for the validity of the large--$q$ Taylor series expansions of the
reduced function $W_{red.}(\{G\},q)$.  In addition to families obtained by
removal of bonds from nonadjacent vertices in the $K_p$ subgraph of
$K_p + G_r$, we have constructed and investigated a number of families of 
graphs by the powerful method of homeomorphic expansion from respective
starting families. We have shown how the families thus obtained have, in the
limit of an infinite number of vertices, noncompact boundaries in the $q$
plane which pass vertically through the origin of the $z=1/q$ plane and have
support for $Re(q) \ge 0$.

This research was supported in part by the NSF grant PHY-97-22101. 

\vspace{20mm}

\section{Appendix 1}

In this Appendix we gather together some convenient formulas concerning the
function $D_k$, defined in eq. (\ref{dk}) in the text.  One has 
\beq
D_k(q=0) = (-1)^k (k-1)
\label{dk0}
\eeq
and
\beq
D_k(q=1)=(-1)^k
\label{dk1}
\eeq
Since for $k$ even, the circuit graph $C_k$ is bipartite, which is equivalent
to the fact that the chromatic number $\chi(C_{k \ even})=2$, and thus 
$P(C_{k \ even},q=2)=2$, it follows that 
\beq
D_{k \ even}(q=2)=1
\label{dkevenq2}
\eeq
Since for $k$ odd, $\chi(C_{k \ odd})=3$ and $P(C_{k \ odd},q=2)=0$, we have 
\beq
D_{k \ odd}(q=2)=0
\label{dkoddq2}
\eeq
This zero results from a linear factor, i.e., 
\beq
D_{k \ odd}(q=2)=(q-2)Pol(q)
\label{dkkoddfactor}
\eeq
where $Pol(q)$ is a polynomial of degree $k-3$ in $q$ with $Pol(q=2) \ne 0$. 
Two identities that we have derived and used for our calculations 
are listed below:
\beq
D_k-aD_{k-1}=(-1)^k
\label{dkdkm1rel}
\eeq
\beq
D_{k+1} - D_k = D_3D_k + (-1)^{k+1}
\label{dkpdkrel}
\eeq
The proofs follow immediately from the definition of $D_k$. 

\pagebreak

\vfill
\eject

\begin{thebibliography}{99}
\bibitem{graphdef}{We recall the strict mathematical definition of a graph as
a finite collection of vertices and bonds connecting (all or a subset of the)
vertices, such that there are (i) no loops, i.e., bonds connecting a
vertex to itself and (ii) no multiple bonds connecting a given pair of 
vertices.   We denote the limit as the number of 
vertices of a family of graphs $G$ goes to infinity as $\{G\}$.}
\bibitem{early}{Relevant earlier papers on ${\cal B}$ are
Refs. \cite{bds}--\cite{read91}.  (The locus ${\cal B}$ for ladder graphs given
in Refs. \cite{bds} and \cite{read91} were not quite correct; the correct 
locus was given in Ref. \cite{w}; see also the comment in Ref. \cite{strip}.)}
\bibitem{exception}{It sometimes happens that a portion of ${\cal B}$ 
is comprised by a line segment $q_{r1} \le q \le q_{r2}$ lying on the positive 
real $q$ axis and, furthermore, that either (i) no part of ${\cal B}$ crosses
the positive real $q$ axis, so no crossing point $q_c$ is defined, or (ii) a
part of ${\cal B}$ does cross the positive real $q$ axis at a value $q_c$, but
$q_c < q_{r2}$, so that, as one moves down the positive real $q$ axis toward
smaller $q$, the first nonanalyticity that $W(\{G\},q)$ has occurs at 
$q_{r2}$ rather than at $q_c$.  From our studies of many 
families of graphs, we have found a few examples where this type of behavior 
occurs, such as for the infinitely long strip of the square lattice of width 
$L_y=4$ vertices ($w=3$ squares) \cite{strip}, an infinitely long strip of the
$(3^3 \cdot 4^2)$ Archimedean lattice of width $L_y=3$ vertices, and a 
homeomorphic expansion of a width $L_y=2$ strip of the triangular lattice.}
\bibitem{complete}{The complete graph $K_n$ on $n$ vertices is defined as the
graph where each of these vertices is connected to all of the other $n-1$
vertices by bonds.  The complement $\overline G$ of a graph $G$ contains the 
same set of vertices such that two vertices are adjacent in $\overline G$ if
and only if they are not adjacent in $G$.  The complement $\overline K_n$ of 
$K_n$ is thus the graph with $n$ vertices and no bonds.}
\bibitem{thm}{We recall the statement of the addition-contraction theorem 
\cite{rtrev}-\cite{biggsbook}: let $G$ be a graph, and let $v$ and $v'$ be two 
non-adjacent vertices in $G$.  Form (i) the graph $G_{add.}$ in which one 
adds a bond connecting $v$ and $v'$, and (ii) the graph $G_{contr.}$ in which 
one identifies $v$ and $v'$. Then the chromatic polynomial for coloring $G$ 
with $q$ colors, $P(G,q)$, satisfies $P(G,q) = P(G_{add.},q) + 
P(G_{contr.},q)$.}
\bibitem{lknote}{In the limit $k \to \infty$ with $r$ fixed (cf. eq. 
(\ref{kinf})) of the families $T_{k,r}$, $C_{k,r}$, and $S_{k,r}$, the 
resultant boundary ${\cal B}$ is given by the compact locus of eq. 
(\ref{blk}).}
\bibitem{mp}{In the technical terminology of algebraic geometry \cite{alg}, 
a multiple
point on an algebraic curve is a point where several branches of the curve
intersect.  If all $n_i$ of the branches have different tangents, the multiple
point is said to have index $n_i$.}
\end{thebibliography}
\end{document}